\documentclass[twocolumn,twocolappendix]{aastex63}
\usepackage{graphicx}
\usepackage{upgreek}

\newcommand{\oii}{[O\,II] }
\newcommand{\oiirm}{\mathrm{[O\,II]}}
\newcommand{\oiii}{[O\,III] }

\newcommand{\M}{M_{\ast}}

\usepackage{amsmath}
\DeclareMathOperator\erf{erf}

\received{XXX}
\revised{YYY}
\accepted{ZZZ}

\submitjournal{ApJ}

\shorttitle{ELG-halo connection}
\shortauthors{Gao et al.}

\begin{document}

\title{Construct the emission line galaxy-host halo connection through auto and cross correlations}

\correspondingauthor{Y.P. Jing}
\email{ypjing@sjtu.edu.cn}

\author{Hongyu Gao}
\affiliation{Department of Astronomy, School of Physics and Astronomy, Shanghai Jiao Tong University, Shanghai 200240, People's Republic of China}
\author[0000-0002-4534-3125]{Y.P. Jing}
\affiliation{Department of Astronomy, School of Physics and Astronomy, Shanghai Jiao Tong University, Shanghai 200240, People's Republic of China}
\affiliation{Tsung-Dao Lee Institute, and Shanghai Key Laboratory for Particle Physics and Cosmology, Shanghai Jiao Tong University, Shanghai 200240, People's Republic of China}
\author{Yun Zheng}
\affiliation{Department of Astronomy, School of Physics and Astronomy, Shanghai Jiao Tong University, Shanghai 200240, People's Republic of China}
\author[0000-0002-7697-3306]{Kun Xu}
\affiliation{Department of Astronomy, School of Physics and Astronomy, Shanghai Jiao Tong University, Shanghai 200240, People's Republic of China}
\begin{abstract}
We investigate the [O\,II] emission line galaxy (ELG)-host halo connection via auto and cross correlations, and propose a concise and effective method to populate ELGs in dark matter halos without assuming a parameterized halo occupation distribution (HOD) model. Using the observational data from VIMOS Public Extragalactic Redshift Survey (VIPERS), we measure the auto and cross correlation functions between ELGs selected by [O\,II] luminosity and normal galaxies selected by stellar mass. Combining the stellar-halo mass relation (SHMR) derived for the normal galaxies and the fraction of ELGs observed in the normal galaxy population, we demonstrate that we can establish an accurate ELG-halo connection. With the ELG-halo connection, we can accurately reproduce the auto and cross correlation functions of ELGs and normal galaxies both in real-space and in redshift-space, once the satellite fraction is properly reduced. Our method provides a novel strategy to generate ELG mock catalogs for ongoing and upcoming galaxy redshift surveys. We also provide a simple description for the HOD of ELGs. 
\end{abstract}

\keywords{Emission line galaxies (459), Redshift surveys (1378), Galaxy dark matter halos (1880), Dark energy (351), Observational cosmology (1146)}

\section{Introduction} \label{sec:intro}
Distinguishing dark energy models from modified gravity theories requires us to accurately measure the entire evolutionary history of the Universe from the matter dominance to the dark energy dominance. By analyzing the clustering of galaxies, we can measure the expansion history and instantaneous expansion rate of the universe from the baryon acoustic oscillation \citep[BAO, e.g.,][]{2005MNRAS.362..505C,2005ApJ...633..560E}, and the growth rate from the redshift-space distortion \citep[RSD, e.g.,][]{1987MNRAS.227....1K}. Sloan Digital Sky Survey \citep[SDSS,][]{2000AJ....120.1579Y, 2006AJ....131.2332G} has performed spectroscopic measurement for a large number of galaxies in the low-redshift Universe that has been dominated by dark energy. But for distant galaxies with redshift $z>0.7$, because their continuum gets faint and most of their optical spectral lines are redshifted to the infrared band, it becomes  more and more difficult to conduct a large redshift survey at high redshift. 

To overcome this difficulty, galaxies with strong \oii emission lines (ELGs) have naturally become the main target for next generation redshift surveys \citep{2013ApJS..208....5N, 2016AJ....151...44D,2016arXiv161100036D,2014PASJ...66R...1T}. Since the neutral gas can be photo-ionized by the ultraviolet (UV) radiation of newly formed massive stars and produce \oii lines, the main population of \oii emitters is therefore expected to be star-forming galaxies. Although the violent nuclear activities caused by super massive black holes (SMBHs) also have enough energy to ionize oxygen atoms, the fraction of active galactic nuclei (AGN) in the \oii emitters is small \citep[e.g.,][]{2013MNRAS.428.1498C}. Compared to other spectral features (such as H$\alpha$, \oiii and $4000 \,\rm{\AA}$ break), the main advantage of the \oii line is the doublet feature at the wavelengths $\lambda 3727, 3729 \, \rm{\AA}$, and it can be detected in the optical window up to redshift 1.6. For example, the Deep Extragalactic Evolutionary Probe 2 \citep[DEEP2,][]{2013ApJS..208....5N} has measured the spectrum for more than 50,000 galaxies at $z \sim 1$, in which the number of \oii ELGs is dominant. Currently, the Dark Energy Spectroscopic Instrument \citep[DESI,][]{2016arXiv161100036D} is conducting spectroscopic observations for more than 17 million \oii ELGs within $0.6<z<1.6$ covering 14,000 square degrees, which makes up for the vacancy of the luminous red galaxy (LRG) sample at $z>1$. With the help of its near-infrared spectrometers and large-aperture, the Subaru Prime Focus Spectrograph \citep[PFS,][]{2014PASJ...66R...1T} will conduct spectroscopic observations of \oii ELGs all the way to $z=2.4$. The combination of these two redshift surveys will increase the number and coverage of observed \oii ELGs to an unprecedented level. 

Before we can extract the cosmological information from the clustering signal of \oii ELGs, we need to understand the connection between these galaxies and the underlying dark matter. This is also a prerequisite for generating realistic \oii ELG mock catalogs \citep[e.g.,][]{2021arXiv210713168O} for these redshift surveys. Halo occupation distribution (HOD) has become one of the most common ways to construct the halo-galaxy connection \citep[e.g.,][]{1998ApJ...494....1J, 2000MNRAS.318.1144P, 2000ApJ...543..503M, 2000MNRAS.318..203S, 2002ApJ...575..587B, 2005ApJ...633..791Z, 2007ApJ...667..760Z, 2011ApJ...736...59Z, 2015MNRAS.454.1161Z, 2016MNRAS.457.4360Z, 2018MNRAS.476.1637Z, 2016MNRAS.459.3040G, 2016MNRAS.460.1173R, 2016MNRAS.460.3647X, 2018MNRAS.481.5470X, 2018MNRAS.478.2019Y, 2019ApJ...879...71W}. Under the HOD framework, the mean occupation number $\left \langle N\left(M\right) \right \rangle$ of a given galaxy population in a halo is determined by the halo mass. In addition to HOD, the statistics related to the physical properties (such as stellar mass and luminosity) of the galaxies inhabiting a halo of given mass can be described by the conditional luminosity (stellar mass) function \citep{2003MNRAS.339.1057Y, 2006MNRAS.365..842C, 2007MNRAS.376..841V, 2009ApJ...695..900Y, 2012ApJ...752...41Y, 2015ApJ...799..130R, 2018ApJ...858...30G}. Since subhalos can be more accurately resolved as cosmological simulations improves, the (sub)halo abundance matching (AM) method \citep[e.g.,][]{1998ApJ...506...19W, 2006MNRAS.371..537W, 2006MNRAS.371.1173V, 2010ApJ...717..379B, 2010MNRAS.402.1796W, 2010MNRAS.404.1111G, 2012MNRAS.423.3458S, 2013MNRAS.428.3121M, 2014MNRAS.437.3228G, 2016MNRAS.459.3040G, 2016MNRAS.460.3100C, 2018ARA&A..56..435W, 2019MNRAS.488.3143B, 2021arXiv210911738X, 2021arXiv211005760X} has become a more effective way to link the observable physical quantity (e.g., stellar mass, luminosity) of a galaxy to its host (sub)halo properties (e.g., halo mass, maximum circular velocity). For normal galaxies in a stellar mass-selected galaxy sample, their galaxy-halo connection is relatively easy to understand, because the monotonically increasing stellar-halo mass relation (SHMR) indicates that the massive galaxies tend to occupy massive halos, although it might be affected by other properties beyond stellar mass due to galaxy assembly bias \citep[e.g.,][]{2010MNRAS.402.1942C, 2013MNRAS.433..515W, 2014MNRAS.443.3044Z, 2015MNRAS.452.1958H, 2016MNRAS.457.3200M, 2020MNRAS.492.2739X, 2020MNRAS.493.5506H, 2021MNRAS.501.1603H, 2021MNRAS.505.5117Z, 2021arXiv210806790Z, 2021NatAs...5.1069C, 2021arXiv211005760X}. However, the situation may become more complicated for ELGs. Since the quench fraction of massive galaxies is relatively higher, galaxies with strong emission lines are not necessarily hosted by massive halos. On the contrary, low-mass galaxies are more likely to have strong star formation processes. Therefore, the probability that a halo hosts an ELG is expected to peak at low-mass, and then gradually decreases toward the high-mass end \citep{2012MNRAS.426..679G, 2013MNRAS.432.2717C}.

Recently, a handful studies have been devoted to studying the \oii ELG-halo connection in observations \citep{2016MNRAS.461.3421F, 2017MNRAS.472..550F, 2019ApJ...871..147G, 2020MNRAS.499.5486A, 2021PASJ...73.1186O}. For instance, by simultaneously constraining the SHMR, completeness and quench fraction of the \oii ELG sample from the extended Baryon Oscillation Spectroscopic Survey \citep[eBOSS,][]{2016AJ....151...44D}, \cite{2019ApJ...871..147G} found that the typical host halo mass of eBOSS ELGs is $\sim 10^{12}\,M_{\odot}$ and the satellite fraction is 13\%-17\%, although the results slightly depend on the assumptions of their quenched fraction model. They also showed that the completeness of eBOSS ELGs is less than 10\%, which indicates that only galaxies with the strongest \oii emissions are selected by eBOSS. Using the \oii ELGs sample identified by the narrow-band (NB) filters at $z>1$ in the
Subaru Hyper-Suprime Cam (HSC) survey, \cite{2021PASJ...73.1186O} found that their angular correlation function can be well fitted by the HOD model proposed by \cite{2012MNRAS.426..679G}, but the constraints of the model parameters are poor due to the limited data and the large parameter space. From the number densities of the HSC NB ELGs ($\sim 10^{-3} \, {\rm Mpc}^{-3}\,h^3$) and eBOSS ELGs ($\sim 10^{-4}\, {\rm Mpc}^{-3}\,h^3$), we can easily see that these ELGs are different populations of \oii emitters. It is important to study how the ELG-halo connection depends on the \oii luminosity. 

Different from previous works, we aim to investigate the ELG-halo connection for different \oii luminosities by utilizing the cross correlations between ELGs and normal galaxies. Since the host halos of ELGs are widely distributed in mass as we will show, the auto correlation of ELGs actually mixes the clustering signal of halos with different mass and is therefore difficult to interpret. But for the normal galaxies selected by stellar mass, we already have a relatively clear understanding of their clustering and host halos properties. Furthermore, the number density of normal galaxies is higher, which makes the cross correlation a better measured quantity. Therefore, the cross correlation of an ELG sample with a stellar mass-selected galaxy sample will tell us how ELGs are distributed around halos which are derived from the SHMR of normal galaxies. In this work, we use the galaxy catalog from the VIMOS
Public Extragalactic Redshift Survey \citep[VIPERS\footnote{http://vipers.inaf.it},][]{2014A&A...566A.108G,2014A&A...562A..23G,2018A&A...609A..84S}. Unlike the eBOSS which only selects ELGs with strong emission lines, VIPERS is an $i$-band limit survey and thus contains ELGs with more moderate \oii luminosity that are also the main targets of DESI and PFS. Different from the HOD modelings mentioned before, we make full use of the ELG-stellar mass relation in observation without establishing a parameterized model. We will demonstrate that by randomly assigning ELGs to dark matter halos according to the SHMR of normal galaxies, we can well repeat the auto and cross correlation functions in both real-space and redshift-space as long as the satellite fraction is reduced. The method is simple but effective, which could be a starting point for constructing the ELG-halo connection for surveys such as DESI and PFS, and become a test-bed for further improving the connection.  

The layout of this paper is organized as follows. In Section \ref{sec:data}, we describe the observational data and numerical simulation used in this work. In Section \ref{sec:clustering}, we introduce our methods to account for the survey selection effects and to measure correlation functions. The SHMR is derived by the AM method in Section \ref{sec:SHMR}. The main results of the ELG-halo connection are presented in Section \ref{sec:populate ELGs}. Eventually, we give a brief conclusion in Section \ref{sec:summary}. Unless otherwise stated, the cosmological parameters used in this paper are: $\Omega_{\mathrm{m},0} = 0.268$, $\Omega_{\Lambda,0} = 0.732$ and $H_0 = 100h \,\mathrm{km\, s^{-1}\,Mpc^{-1}}=71 \,\mathrm{km\,s^{-1}\,Mpc^{-1}}$.

\begin{figure*}
	\centering
	\includegraphics[scale=0.8]{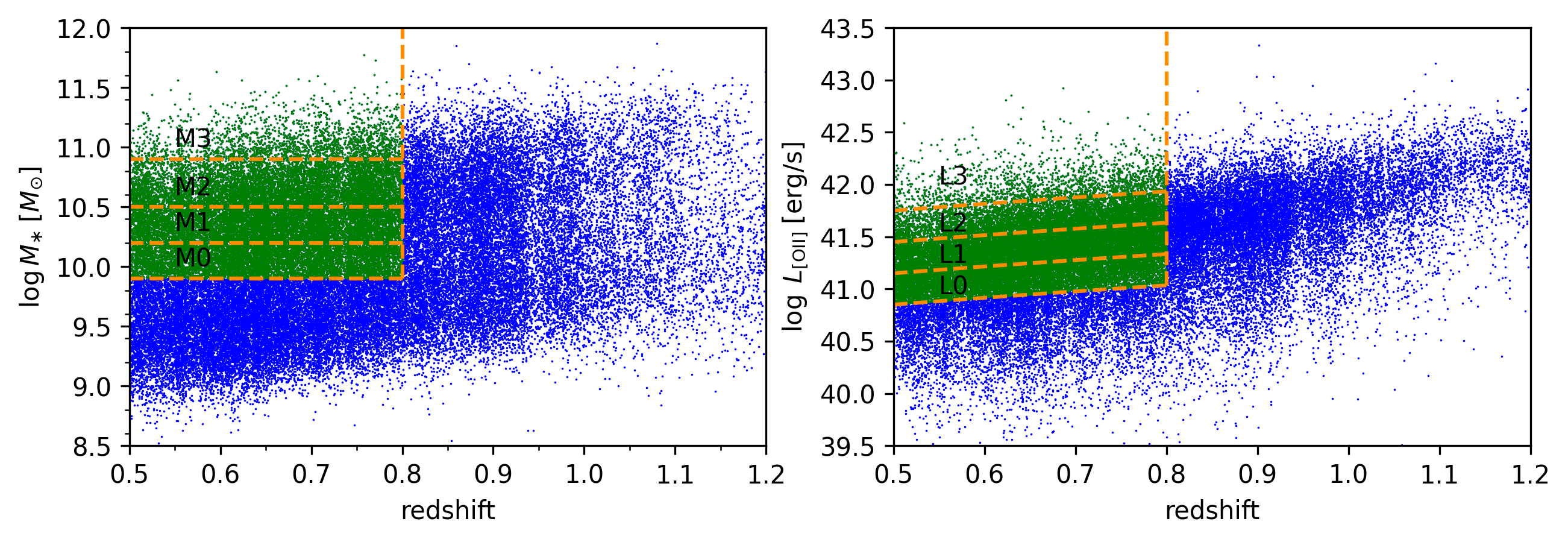}
	\caption{The stellar mass and \oii luminosity distributions of galaxies in the VIPERS sample. The blue points denote the full galaxy sample while the green points represent the subsamples, whose boundaries are plotted as orange dashed lines. 
		\label{fig:SM_and_LOII_z_05_dz03}}
\end{figure*}

\begin{deluxetable}{cccc}
	\tablenum{1}
	\tablecaption{The details of four stellar mass-selected subsamples. \label{tab:stellar mass-selected}}
	\tablehead{ \colhead{Name} &
		\colhead{Redshift Range} & \colhead{$\log \M\, [M_{\odot}]$ } &\colhead{$N_{\mathrm{g}}$}
	}
	\startdata
	$M0$&$0.5<z\leq 0.8$& $\left[9.9, 10.2\right]$& 7829 \\
	$M1$&$0.5<z\leq 0.8$& $\left[10.2, 10.5\right]$& 8355 \\
	$M2$&$0.5<z\leq 0.8$& $\left[10.5, 10.9\right]$& 8040 \\
	$M3$&$0.5<z\leq 0.8$& $\left[10.9, \infty\right]$& 1964 \\
	\enddata
\end{deluxetable}

\begin{deluxetable}{cccc}
	\tablenum{2}
	\tablecaption{The details of four \oii luminosity-selected subsamples. \label{tab:luminosity-selected}}
	\tablehead{\colhead{Name} &
		\colhead{Redshift Range} & \colhead{$\log L_\oiirm \left(z=0.5\right) \, \mathrm{[erg/s]}$ } &\colhead{$N_{\mathrm{g}}$}
	}
	\startdata
	$L0$&$0.5<z\leq 0.8$& $\left[40.85, 41.15\right]$& 9349 \\
	$L1$&$0.5<z\leq 0.8$& $\left[41.15, 41.45\right]$& 11721 \\
	$L2$&$0.5<z\leq 0.8$& $\left[41.45, 41.75\right]$& 6281 \\
	$L3$&$0.5<z\leq 0.8$& $\left[41.75, \infty\right]$& 1693 \\
	\enddata
\end{deluxetable}

\begin{deluxetable}{ccccc}
	\tablenum{3}
	\tablecaption{The fractions of $L_\oiirm$-selected galaxies included in each $\M$-selected subsample. \label{tab:fraction}}
	\tablehead{ \colhead{Name} &
		\colhead{$L0$ fraction} & \colhead{$L1$ fraction} & \colhead{$L2$ fraction} & \colhead{$L3$ fraction} 
	}
	\startdata
	$M0$&0.210&0.243&0.120&0.039 \\
	$M1$&0.181&0.134&0.065&0.020 \\
	$M2$&0.172&0.108&0.043&0.016 \\
	$M3$&0.149&0.108&0.045&0.024 \\
	\enddata
\end{deluxetable}

\section{Galaxy sample and simulation data} \label{sec:data}
We describe the basics of VIPERS and the properties of our \oii luminosity-selected and stellar mass-selected subsamples. The N-body cosmological simulation used in this study is also introduced in this Section.

\subsection{VIPERS sample} \label{subsec:galaxy sample}
We use the galaxy catalog of the final public release (PDR-2) \citep{2018A&A...609A..84S} of the VIPERS. This survey overlaps two sky fields W1 and W4 of the Canada-France-Hawaii Telescope Legacy Survey Wide (CFHTLS-Wide\footnote{http://www.cfht.hawaii.edu/Science/CFHLS/}), covering about 24 square degrees. The multi-band magnitudes ($u, g, r, i, z$) of the parent photometric catalog come from the CFHTLS T0005\footnote{http://www.cfht.hawaii.edu/Science/CFHLS/T0005/}. Ancillary photometric data is supplemented by the VIPERS Multi-Lambda Survey \citep{2016A&A...590A.102M}, which matched the CFHTLS T0005 catalog with GALEX \citep{2005ApJ...619L...1M} and the VISTA Deep Extragalactic Observations \citep{2013MNRAS.428.1281J}, and provided extra photometry in NUV, FUV and $K_{\mathrm{s}}$ $(K_{\mathrm{video}})$ bands. Galaxies with $i_{\mathrm{AB}}<22.5$ in the parent catalog satisfying the following color-color criteria
\begin{eqnarray}
\left(r-i\right)>0.5\times\left(u-g\right) \quad \mathrm{OR} \quad \left(r-i\right)>0.7  \label{equ:color-cut}
\end{eqnarray}
are selected as the spectroscopic targets. The spectra of about 90,000 galaxies were measured with the VIMOS multi-object spectrograph on the ESO Very Large Telescopes \citep{2003SPIE.4841.1670L}. Finally, we only include the VIPERS main galaxy targets ($\mathrm{classFlag}=1$) with high-quality redshift measurements ($\mathrm{zflag}\geq2$) in our research.
\subsection{galaxy subsamples} \label{subsec:galaxy subsample}
For the purpose of analyzing the cross correlations of emission line galaxies and normal galaxies, we divide the galaxy sample in the redshift range $0.5<z\leq0.8$ into four \oii luminosity $L_\oiirm$-selected ($L0$, $L1$, $L2$ and $L3$) and four stellar mass $\M$-selected ($M0$, $M1$, $M2$ and $M3$) subsamples. After subtracting the continuum, the \oii fluxes are measured by fitting a single Gaussian model to the spectrum. The Levenberg-Marquardt algorithm \citep{levenberg1944method, marquardt1963algorithm} is adopted to derive the best-fitting \oii fluxes and their uncertainties. We take the multi-band photometry from VIPERS Multi-Lambda Survey \citep{2016A&A...590A.102M} to model the spectral energy distribution (SED) of galaxies. The {\tt\string LE PHARE} \citep{2002MNRAS.329..355A,2006A&A...457..841I} code is used to perform the SED fitting and estimate the physical properties (including stellar mass) of galaxies. More details about the $L_\oiirm$ measurements and SED template settings can be found in \cite{2021ApJ...908...43G}\footnote{The cosmological parameters used in the \oii luminosity computation and SED fitting process are $\Omega_{\Lambda,0} = 0.7$, $\Omega_{m,0} = 0.3$ and $H_0 = 70\,\mathrm{km \, s^{-1} \, Mpc^{-1}}$, which are slightly different from what we adopted in this study. But this does not affect our subsequent analysis, because we mainly care about the relative difference between different subsamples rather than their absolute value.}.

At $0.5<z\leq0.8$, there are a total of 45,600 galaxies, of which 36,741 have $L_\oiirm>0$. The mass and \oii luminosity distributions of the full sample are displayed as blue points in Figure \ref{fig:SM_and_LOII_z_05_dz03}, in which the four $\M$-selected and $L_\oiirm$-selected subsamples are also shown as green points. We present more details of each subsample in Table \ref{tab:stellar mass-selected} and \ref{tab:luminosity-selected}.  Considering that a galaxy may be contained in both a $\M$-selected and a $L_\oiirm$-selected subsample, we present the fraction of $L_\oiirm$-selected galaxies included in each $\M$-selected subsample in Table \ref{tab:fraction}. This fraction represents the degree of independence of the two subsamples.

Since the stellar mass function (SMF) evolves relatively weakly at $z<1$ \citep{2007A&A...474..443P, 2010A&A...523A..13P, 2013A&A...558A..23D} , we apply flat stellar mass cuts to construct $\M$-selected subsamples. In order to determine the stellar mass completeness limit of the galaxy sample, we follow the same technique proposed by \cite{2010A&A...523A..13P} (see also \cite{2013A&A...558A..23D}). The 90\% stellar mass completeness limit at $z \sim 0.6$ is $10^{9.8} \, M_{\odot}$, so we take a lower boundary $\M  = 10^{9.9} \, M_{\odot}$  for the first subsample $M0$.

As for the $L_\oiirm$-selected subsamples, we adopt redshift evolution cuts to account for the evolution of \oii luminosity function. Referring to the parameterized evolution model of the characteristic luminosity $L_{\mathrm{\oii},\star}\left(z\right)=L_{\mathrm{\oii},\star}\left(0\right)\left(1+z\right)^{\beta_L}$ provided by \cite{2016MNRAS.461.1076C}, we define the $L_\oiirm$ cut at $z$ as 
\begin{eqnarray}
\log L^{\mathrm{cut}}_\oiirm \left(z\right) = \log L^{\mathrm{cut}}_\oiirm \left(z=0.5\right) + \log \left(\frac{1+z}{1+0.5}\right)^{\beta_L}
\end{eqnarray}
with $\beta_L=2.33$. To ensure the completeness of \oii detection, we set the lower boundary of the first subsample $L0$ as $ L_\oiirm \left(z=0.5\right)=10^{40.85} \, \mathrm{erg\,s^{-1}}$, which is above the $5\sigma$ detection limit ($F^{\mathrm{lim}}_\oiirm = 3.5\times 10^{-17} \, \mathrm{erg\,s^{-1}\,cm^{-2}}$) of the VIMOS with an exposure time of 0.75 hours \citep{2016MNRAS.461.1076C}. 

\begin{figure}
	\centering
	\includegraphics[scale=0.8]{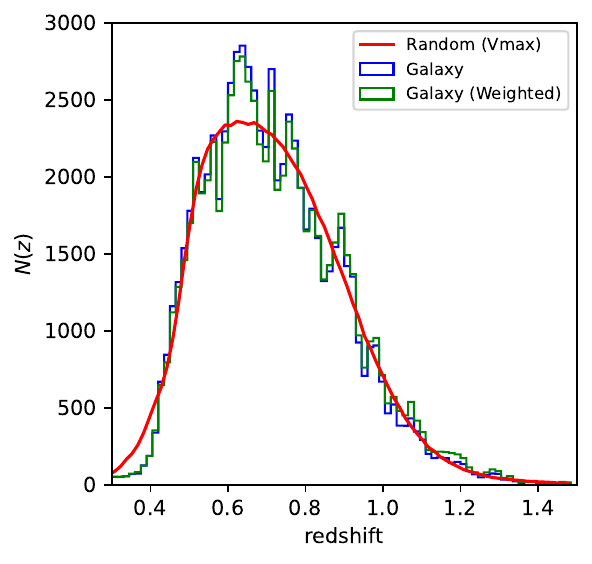}
	\caption{The redshift distributions of the galaxy and random samples. The blue histogram shows the full galaxy sample (W1 and W4 fields) of VIPERS. The distribution of the galaxy sample weighted by $w^{\mathrm{SSR}}\times w^{\mathrm{TSR}}$ is displayed as the green histogram, which shows a slightly difference from the un-weighted one. The red solid line represents the radial distribution of the random sample generated with the $V_{\mathrm{max}}$ technique as described in the text. The numbers of weighted galaxies and random points have been properly normalized. 
		\label{fig:redshift_distribution}}
\end{figure}

\subsection{Simulation} \label{subsec:simulation}
One of the high resolution N-body cosmological simulations of CosmicGrowth \citep{2019SCPMA..6219511J} is used to study the galaxy-halo connection. This simulation is performed by the $\mathrm{P^3M}$ method \citep{2002ApJ...574..538J} and has $3072^3$ dark matter particles in a $600\, \mathrm{Mpc}\,h^{-1}$ box with the standard $\Lambda$CDM cosmological parameters: $\Omega_{\mathrm{m}} = 0.268$, $\Omega_{\Lambda} = 0.732$, $h=0.71$, $n_{\mathrm{s}}=0.968$ and $\sigma_8=0.83$. The halos and subhalos are identified by the friends-of-friends algorithm (FOF) \citep{1985ApJ...292..371D} and the Hierarchical-Bound-Tracing algorithm (HBT+) \citep{2012MNRAS.427.2437H,2018MNRAS.474..604H}, respectively.

Although the \oii ELGs are considered to be more likely to exist in low-mass halos ($\sim 10^{12} \, M_{\odot}$) \citep[e.g.,][]{2016MNRAS.461.3421F, 2019ApJ...871..147G, 2021MNRAS.502.3599H, 2021PASJ...73.1186O}, the mass resolution $m_{\mathrm{p}} = 5.54 \times 10^8 \,M_{\odot}\,h^{-1}$ of particles in our simulation is sufficient to resolve them. The halo mass $M_{\mathrm{h}}$ is defined as its viral mass $M_{\mathrm{vir}}$ that is the mass enclosed by a sphere with an average density of $\Delta_{\mathrm{vir}}\left(z\right)$ times the critical density of the universe \citep{1972ApJ...176....1G,1998ApJ...495...80B}. The subhalo accretion mass $M_{\mathrm{s}}$ is defined as its virial mass at the last snapshot before merging into the current host halo. In addition, we have carefully dealt with the small subhalos that have been almost or even completely stripped by the tidal force. Using the fitting formula proposed by \cite{2008ApJ...675.1095J}, we trace the merger history of those subhalos with less than 20 particles and calculate their merger time scale to judge whether they can remain distinct as subhalos. Finally, the snapshot with $z=0.663$, which is close to the mean redshift of our galaxy sample, is chosen for our analysis. To make a fair comparison with the observations, we have incorporated the RSD effects to these simulated halos (subhalos). We choose the $z$-axis as the line of sight and define the redshift of the center of the simulation box as $0.663$. For a halo (subhalo), the cosmological redshift $z_\mathrm{c}$ is given according to its comoving distance to the center of the box, while the redshift $z_\mathrm{p}$ caused by peculiar motion is calculated by $v_{z}/c$, where $v_{z}$ is the velocity in z-direction and $c$ is the speed of light. Besides, we add  to $v_{z}$ a velocity randomly derived from a Gaussian distribution of the dispersion $\sigma_v=c\sigma_z$, where $\sigma_z=0.00054$ is the typical redshift uncertainty for VIPERS \citep{2018A&A...609A..84S}. Finally, the updated $z$-axis coordinate of a halo (subhalo) is converted from its observed redshift $z_\mathrm{obs}=\left(1+z_\mathrm{c}\right) \left(1+z_\mathrm{p}\right) - 1$.  

\section{Measurement of galaxy clustering} \label{sec:clustering}
In this Section, we carefully correct the selection effects in VIPERS and measure the cross (auto) correlation functions for different galaxy subsamples both in redshift-space and real-space.
\subsection{Selection functions} \label{subsec:selection function}
In order to achieve an accurate measurement of galaxy clustering, we should understand and correct for the selection functions listed below.

\begin{enumerate}
	\item The survey masks in VIPERS. We can account for these survey masks by applying the same sky geometry to the random sample.
    \item Target sampling rate (TSR). Some galaxies in the parent photometric catalog cannot be spectroscopically observed due to the limited number of slits. This effect can bias the targeting of galaxies in the dense region due to the uniform distribution of the slits, and the clustering of galaxies is underestimated. This effect can be corrected by up-weighting $w^{\mathrm{TSR}}=1/\mathrm{TSR}$ \citep{2013A&A...557A..54D, 2018A&A...609A..84S} for each galaxy.
    \item Spectroscopic success rate (SSR). It quantifies the probability that the redshift of a galaxy targeted by the VIMOS can be successfully measured (i.e. $\mathrm{zflag}\geq2$). By exploring the dependence of SSR on the multi-dimensional parameter space, \cite{2018A&A...609A..84S} evaluates the SSR for each galaxy based on a nearest-neighbor
    algorithm. Therefore we also up-weight each galaxy with $w^{\mathrm{SSR}}=1/\mathrm{SSR}$.
    \item Slit collisions. Similar to fiber collision, if the distance between the two galaxies is less than the physical size of the silt, only the spectrum of one galaxy can be observed. Additionally, in order to avoid the overlap of spectra in the VIMOS detectors, the spectra of two galaxies with distance below a specific size along the direction perpendicular to silt cannot be observed at the same time. The combination of these two effects will suppress the clustering of galaxies at small scales. Follow the method of \cite{2017A&A...604A..33P}, we calculate the angular weights $w^{\mathrm{A}}\left(\theta\right)$ using 153 VIPERS mock samples (see APPENDIX \ref{sec:slit collisions}) to correct the slit collision effect for galaxy pairs. 
    \item $i$-band magnitude limit and color sampling rate (CSR). VIPERS adopts the $i^{\mathrm{AB}}<22.5$ flux cut and a color-color cut (Equation \ref{equ:color-cut}) to construct a flux-limited sample at $z>0.5$, which introduces two radial selection functions to the redshift distribution of galaxies. \cite{2014A&A...566A.108G} provides an accurate model $\mathrm{CSR}\left(z\right)=1/2 -1/2 \erf\left[b\left(z_{\mathrm{t}}-z\right)\right]$ with $b=10.8$ and $z_{\mathrm{t}}=0.444$ for the radial weight $w^{\mathrm{CSR}}=1/\mathrm{CSR}$ to describe the completeness of the color-color selection. To account for the combination of these two radial selection effects, we use the $V_{\mathrm{max}}$ method \citep{2011MNRAS.416..739C, 2013A&A...557A..54D, 2017A&A...604A..33P, 2017A&A...608A..44D, 2020RAA....20...54Y} to generate a smooth redshift distribution for the random sample. We present the detail of the $V_{\mathrm{max}}$ method in APPENDIX \ref{sec:vmax}. In Figure \ref{fig:redshift_distribution}, we can see that the redshift distribution of the random sample thus generated  is well consistent with the observed one. 
\end{enumerate}
After considering these selection effects, we can estimate the completeness-corrected number density of our galaxy subsamples through
\begin{eqnarray}
n_{\mathrm{g}} = \sum_{i}^{N_{\mathrm{g}}}\frac{w^{\mathrm{TSR}}_i w^{\mathrm{SSR}}_i w^{\mathrm{CSR}}_i}{V_{\mathrm{max},i}}, \label{equ:number density}
\end{eqnarray}
where $V_{\mathrm{max},i}$ is computed with 
\begin{eqnarray}
V_{\mathrm{max},i}=\frac{A_{\mathrm{eff}}}{3\times\left(180/\pi\right)^2}\times  \left[D^3_{\mathrm{com}}\left(z_{\mathrm{max},i}\right) - D^3_{\mathrm{com}}\left(0.5\right) \right],
\end{eqnarray}
where $A_{\mathrm{eff}}=16.324 \, \mathrm{deg^2}$ is the effective sky area of VIPERS and $D_{\mathrm{com}}$ is the comoving distance.

\subsection{Estimation of correlation function} \label{subsec:Estimation of correlation function}             
To measure the galaxy clustering in redshift-space, we decompose the separation vector $\boldsymbol{s}=\boldsymbol{s_1}-\boldsymbol{s_2}$ of two galaxies into two components $r_{\pi}$ and $r_{\mathrm{p}}$. $r_{\pi}$ can be obtained by projecting $\boldsymbol{s}$ along the line-of-sight, $r_{\pi}=\left( \boldsymbol{s} \cdot \boldsymbol{l} \right)/\left| \boldsymbol{l} \right|$ with $\boldsymbol{l}=\left(\boldsymbol{s_1}+\boldsymbol{s_2}\right)/2$, and $r_{\mathrm{p}}$ is calculated as $\sqrt{s^2-r^2_{\pi}}$. We choose twelve $r_{\mathrm{p}}$ bins from $0.12$ to $30 \,\mathrm{Mpc}\,h^{-1}$ with an equal logarithmic interval and forty $r_{\pi}$ bins from $0$ to $40\, \mathrm{Mpc}\,h^{-1}$ with an equal linear interval. The redshift-space cross (auto) correlation functions for different galaxy subsamples are measured utilizing the Landy-Szalay estimator \citep{1993ApJ...412...64L, 1998ApJ...494L..41S}
\begin{eqnarray}
\xi_{xy}\left(r_{\mathrm{p}}, r_{\mathrm{\pi}}\right) = \left[ \frac{D_xD_y-D_xR_y-D_yR_x+R_xR_y}{R_xR_y}\right],
\end{eqnarray}
where $x$, $y$ indicate different samples ($x=y$ for the auto correlation). The normalized pair counts for galaxy-galaxy, galaxy-random and random-random are calculated by 
\begin{eqnarray}
D_xD_y\left(r_{\mathrm{p}}, r_{\pi}\right)&=&\frac{\sum_{i=1}^{N_{\mathrm{g},x}}\sum_{j=1}^{N_{\mathrm{g},y}}w^{\mathrm{A}}\left(\theta_{ij}\right)w^{\mathrm{c}}_{i}w^{\mathrm{c}}_{j}\Theta_{ij}\left(r_{\mathrm{p}}, r_{\pi}\right)}{\sum_{i=1}^{N_{\mathrm{g},x}}\sum_{j=1}^{N_{\mathrm{g},y}}w^{\mathrm{A}}\left(\theta_{ij}\right)w^{\mathrm{c}}_{i}w^{\mathrm{c}}_{j}} \nonumber \\
D_xR_y\left(r_{\mathrm{p}}, r_{\pi}\right)&=&\frac{\sum_{i=1}^{N_{\mathrm{g},x}}\sum_{j=1}^{N_{\mathrm{r},y}}w^{\mathrm{c}}_{i}\Theta_{ij}\left(r_{\mathrm{p}}, r_{\pi}\right)}{N_{\mathrm{r},y} \sum_{i=1}^{N_{\mathrm{g},x}} w^{\mathrm{c}}_{i}} \nonumber \\
D_yR_x\left(r_{\mathrm{p}}, r_{\pi}\right)&=&\frac{\sum_{i=1}^{N_{\mathrm{g},y}}\sum_{j=1}^{N_{\mathrm{r},x}}w^{\mathrm{c}}_{i}\Theta_{ij}\left(r_{\mathrm{p}}, r_{\pi}\right)}{N_{\mathrm{r},x} \sum_{i=1}^{N_{\mathrm{g,y}}} w^{\mathrm{c}}_{i}} \nonumber \\
R_xR_y\left(r_{\mathrm{p}}, r_{\pi}\right)&=&\frac{\sum_{i=1}^{N_{\mathrm{r},x}}\sum_{j=1}^{N_{\mathrm{r},y}}\Theta_{ij}\left(r_{\mathrm{p}}, r_{\pi}\right)}{N_{\mathrm{r},x}N_{\mathrm{r},y}},
\end{eqnarray} 
where $\Theta_{ij} $ is equal to 1 only when a galaxy pair falls into this $\left(r_{\mathrm{p}}, r_{\pi}\right)$ bin, and the pair counts have been up-weighted by $w^{\mathrm{c}}=w^{\mathrm{SSR}}\times w^{\mathrm{TSR}}$ and $w^{\mathrm{A}}\left(\theta \right)$ as mentioned in Section \ref{subsec:selection function}. 

The $\xi_{xy} \left(r_{\mathrm{p}}, r_{\mathrm{\pi}}\right)$ is integrated along the line-of-sight to give the real-space projected correlation function \citep{1983ApJ...267..465D}
\begin{eqnarray}
w^{xy}_{\mathrm{p}}\left(r_{\rm p}\right) = 2\int_{0}^{r_{\pi, \rm max}} \xi_{xy} \left( r_{\rm p}, r_{\pi} \right)\mathrm{d}r_{\pi},
\end{eqnarray}
with $r_{\pi, \rm max}=40 \,\mathrm{Mpc}\,h^{-1}$. We also employ the same upper limit of the integration when modeling the $w^{xy}_{\mathrm{p}}\left(r_{\rm p}\right)$ in the simulation to make a fair comparison.

The covariance matrix of the measured $w^{xy}_{\mathrm{p}}\left(r_{\rm p}\right)$ is estimated with the jackknife technique. We divide the entire survey into 24 fields (16 for W1 and 8 for W4) with an area of approximately one square degree for each field, and the covariance matrix of the measured $w^{xy}_{\mathrm{p}}\left(r_{\rm p}\right)$ can be estimated with 
\begin{eqnarray}
C \left( i,j\right) = \frac{N_{\mathrm{jack}}-1}{N_{\mathrm{jack}}} \sum_{k=1}^{N_{\mathrm{jack}}} \left( w^k_{\mathrm{p}, i} - \bar{w}_{\mathrm{p}, i} \right) \left( w^k_{\mathrm{p}, j} - \bar{w}_{\mathrm{p}, j} \right), \label{equ:covariance}
\end{eqnarray}
where $N_{\mathrm{jack}}=24$ is the number of jackknife samples and $i$ ($j$) denotes the $i$ ($j$)-th $r_\mathrm{p}$ bin.

In analogy to the way of $w_{\mathrm{p}}\left(r_{\rm p}\right)$ in real-space, we also measure the multiple moments of the correlation functions in redshift-space. The monopole $\xi_0\left(s\right)$, quadrupole $\xi_2\left(s\right)$ and hexadecapole $\xi_4\left(s\right)$ \citep{1992ApJ...385L...5H} are defined as
\begin{eqnarray}
\xi_l\left(s\right)=\frac{2l+1}{2}\int_{-1}^{1}\xi\left(s,\mu\right)L_l\left(\mu\right)
\end{eqnarray}
where $L_l\left(\mu\right)$ is the Legendre function, $s$ is binned from $0.12$ to $30 \,\mathrm{Mpc}\,h^{-1}$ with an equal logarithmic interval and the $\mu$ is binned with a linear width $\Delta \mu = 0.1$.

\begin{figure*}
	\centering
	\includegraphics[scale=0.55]{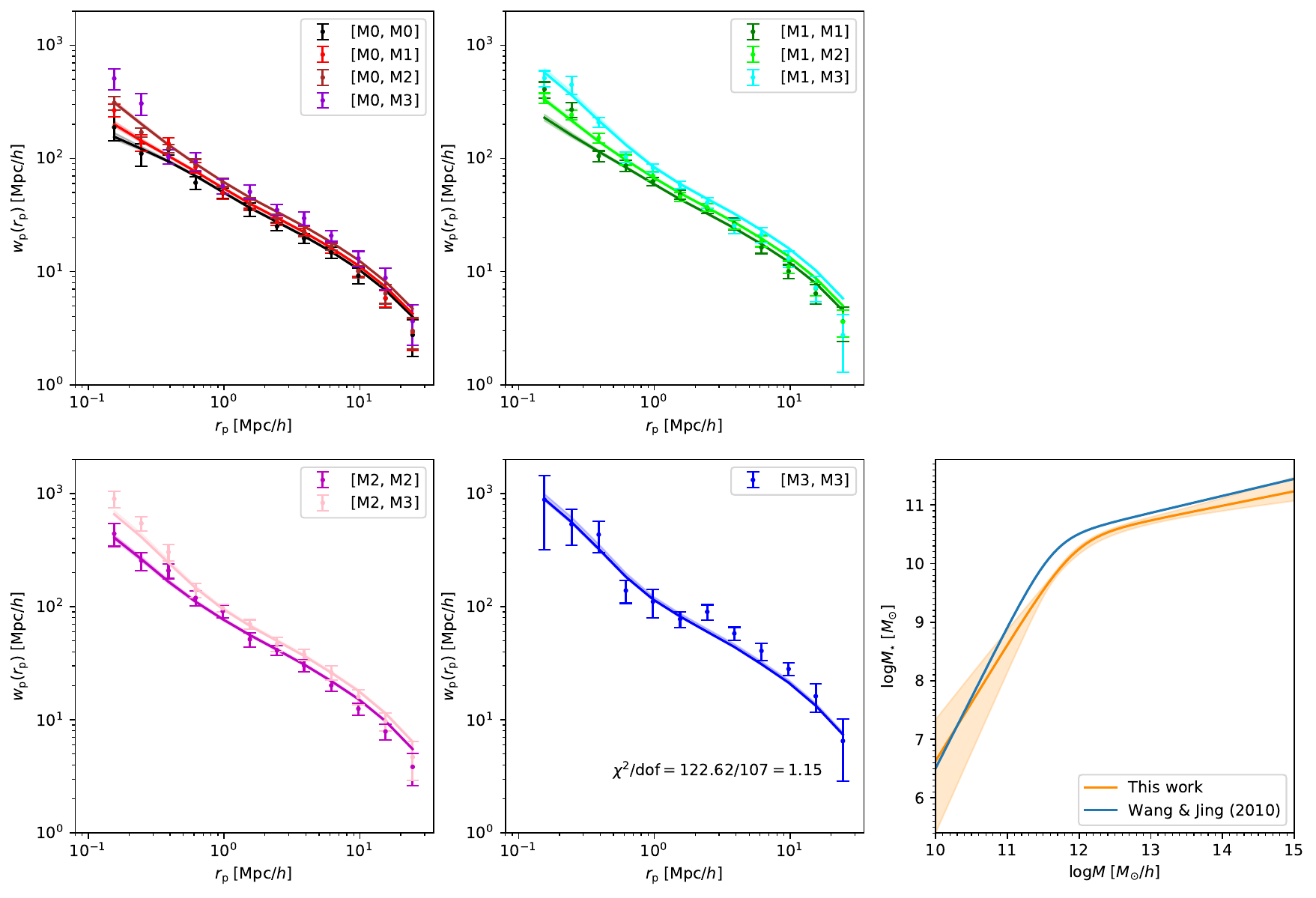}
	\caption{The projected cross (auto) correlation functions for the $\M$-selected subsamples, and the SHMR model derived by the AM approach. In the left four panels, the data points with error bars denote the observational measurements. The $\boldsymbol{w}_{\mathrm{p}}$ between different subsamples is marked with different colors. Except for $\boldsymbol{w}^{M0M3}_{\mathrm{p}}$ of $[M0,M3]$, all observation data are used in the fitting process (see Section \ref{Fitting procedure} for details). The best-fitting $\boldsymbol{w}_{\mathrm{p}}$ model as well as its $1\sigma$ scatter is plotted as solid line with shadow region. The reduced $\chi^2$ is also denoted in the fourth panel. We present the best-fitting SHMR model in the rightmost panel. The SHMR model derived by \cite{2010MNRAS.402.1796W} is also shown as the blue solid line.
		\label{fig:bestfitting_cross_mcmc_SameCS_9SM_table_with_sigmaz_wp_ZMAX_SM_omegam0268}}
\end{figure*}

All the measured correlation functions are shown as data points with error bars in Figures \ref{fig:bestfitting_cross_mcmc_SameCS_9SM_table_with_sigmaz_wp_ZMAX_SM_omegam0268}, \ref{fig:nofsat_sim_cross_randomSM_SameCS_9SM_table_with_sigmaz_wp_ZMAX_LOII_omegam0268.pdf}, \ref{fig:bestfitting_sim_cross_randomSM_SameCS_9SM_table_with_sigmaz_wp_ZMAX_LOII_omegam0268} and \ref{fig:bestfitting_sim_cross_randomSM_SameCS_9SM_table_with_sigmaz_xi024shift_ZMAX_LOII_omegam0268}.

\begin{figure*}
	\centering
	\includegraphics[scale=0.5]{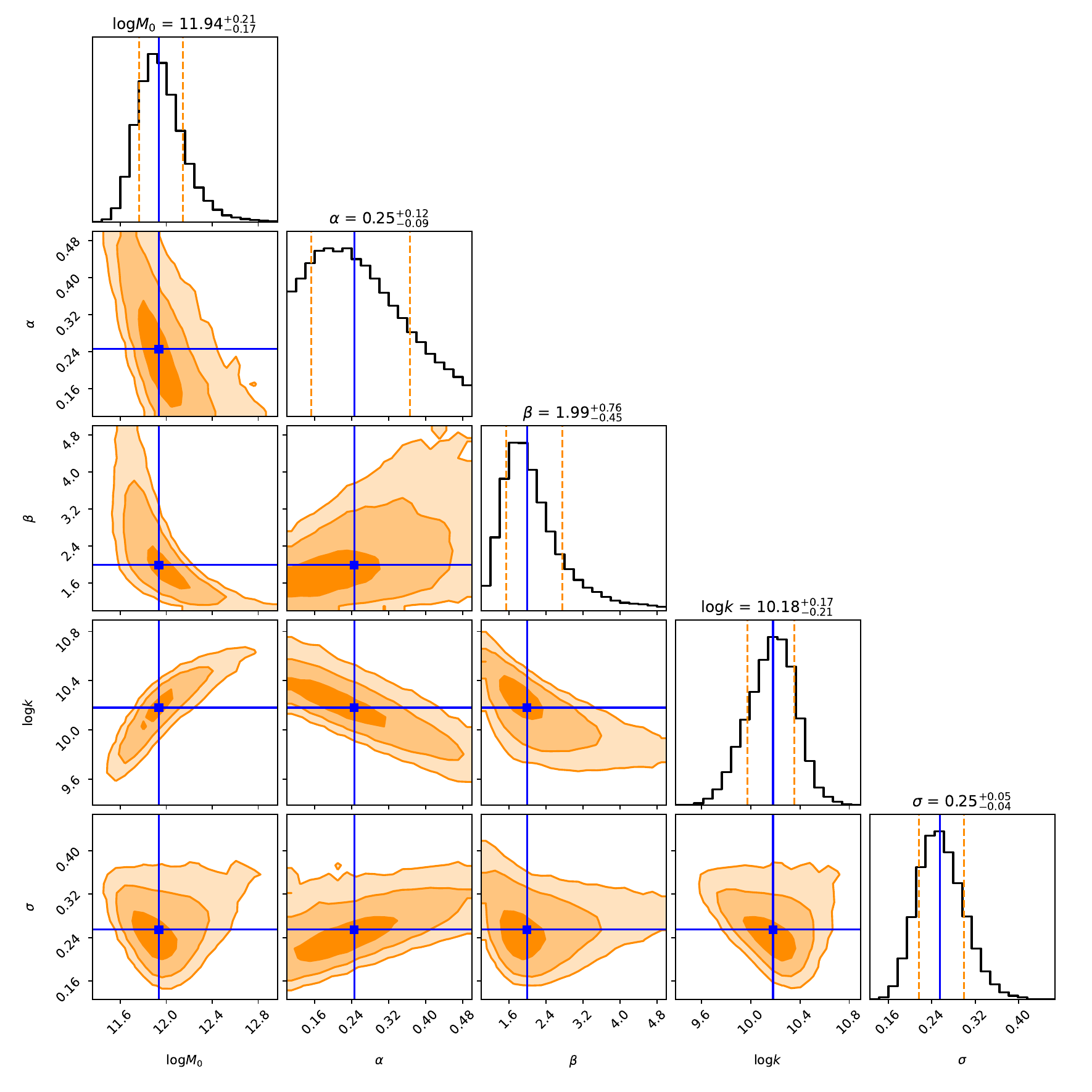}
	\caption{The posterior distributions of the parameters in the unified SHMR model. The 1-D PDF of each parameter is plotted as a histogram at the top panel of each column, where the median value and $1\sigma$ uncertainty is also labeled. The 2-D joint PDF of each parameter pair is shown as a contour with three confidence levels ($68\%$, $95\%$ and $99\%$).   
		\label{fig:fit_corner_cross_mcmc_SameCS_9SM_table_with_sigmaz_wp_ZMAX_SM_omegam0268}}
\end{figure*}

\section{Determining the stellar-halo mass relation} \label{sec:SHMR}
Before modeling the cross correlations of ELGs with normal galaxies, we first establish the connection between stellar mass of normal galaxies and their halo mass. To derive the SHMR by AM method, we use the model proposed by \cite{2010MNRAS.402.1796W} and adopt an efficient way \citep{2016MNRAS.458.4015Z, 2016MNRAS.459.3040G} to calculate the modeled correlation functions. 

\subsection{The abundance matching model} \label{sec:the The abundance matching model}
The conditional probability distribution function (PDF) that a galaxy with stellar mass $\M$ is hosted by a (sub)halo with mass $M$ is assumed to obey a Gaussian distribution
\begin{eqnarray}
p(\M|M) = \frac{1}{\sqrt{2\pi}\sigma}\exp\left[-\frac{\left(\log \M - \log \left\langle \M|M \right\rangle\right)^2}{2\sigma^2}\right]. \label{equ:p(M_star|M)}
\end{eqnarray}
We adopt the parameterized mean relation $\left\langle \M|M \right\rangle$ proposed by \cite{2010MNRAS.402.1796W} (see also \cite{2006MNRAS.371..537W})
\begin{eqnarray}
\left\langle \M|M \right\rangle = \frac{2k}{\left(\frac{M}{M_0}\right)^{-\alpha} + \left(\frac{M}{M_0}\right)^{-\beta}},
\end{eqnarray}
where $\alpha$ and $\beta$ quantify the slopes of two power-law forms separated at $M_0$, and $k$ is a normalization constant. In principle, the $p(\M|M)$ for central and satellite galaxies should be modeled separately to account for possible different formation histories. For the satellites, the current stellar mass depends not only on the accretion mass but also on the subsequent evolution after infalling \citep{2012ApJ...752...41Y}. However, the difference of the $\M$-$M$ relationship between halo and subhalo is small \citep{2010MNRAS.402.1796W}, and the difference should not be important given the current sample size of VIPERS (see below). Therefore, we adopt a unified $\M$-$M$ relationship for halos and for subhalos with the same set of parameters: $\alpha, \beta, M_0, k, \sigma$.
\subsection{The tabulated correlation functions} \label{sec:The tabulated correlation functions}
When exploring the parameter space, we usually need to populate halos (subhalos) with galaxies based on updated AM model parameters and calculate the correlation functions of modeled galaxies by many times. It will consume a significant amount of CPU time if the correlation functions are not calculated efficiently. Therefore, we extend the tabulated method \citep{2016MNRAS.458.4015Z, 2016MNRAS.459.3040G} to calculate the cross correlation function in the simulation. The key of this method is to prepare a table for the correlation functions of different halos (subhalos) binned by mass or other physical properties. Different weights are assigned to the tabulated correlation functions according to the AM model, and the combination yields the correlation function of the modeled galaxies. In this way, the halos (subhalos) in our simulation are divided into 500 tiny mass bins with a width of $\Delta \log M = 0.01$ ranging from $10^{10}$ to $10^{15}\,M_{\odot}\,h^{-1}$. The correlation functions of halo-halo, halo-subhalo and subhalo-subhalo for these bins are then measured and organized into three tables each with $500 \times 500$ elements. Eventually, the modeled $w^{xy}_{\mathrm{p,m}}$ for two galaxy samples $x$ and $y$ is computed by
\begin{eqnarray}
\begin{aligned}
w^{xy}_{\mathrm{p,m}} (r_{\mathrm{p}})  = \\ \sum_{i,j}\frac{\bar{n}_{\mathrm{h},i} \bar{n}_{\mathrm{h},j} }{n^{x}_{\mathrm{g,m}}   n^{y}_{\mathrm{g,m}}}&P_x\left(M_{\mathrm{h},i}\right) P_y\left(M_{\mathrm{h},j}\right) w_{\mathrm{p,hh}}\left(r_{\mathrm{p}}|M_{\mathrm{h},i},M_{\mathrm{h},j}\right)\\
+ \sum_{i,j}\frac{\bar{n}_{\mathrm{h},i} \bar{n}_{\mathrm{s},j} }{n^{x}_{\mathrm{g,m}} n^{y}_{\mathrm{g,m}}} [&P_x\left(M_{\mathrm{h},i}\right) P_y\left(M_{\mathrm{s},j}\right) \\
+ &P_y\left(M_{\mathrm{h},i}\right) P_x\left(M_{\mathrm{s},j}\right)] w_{\mathrm{p,hs}}\left(r_{\mathrm{p}}|M_{\mathrm{h},i},M_{\mathrm{s},j}\right)\\
+\sum_{i,j}\frac{\bar{n}_{\mathrm{s},i} \bar{n}_{\mathrm{s},j} }{n^{x}_{\mathrm{g,m}} n^{y}_{\mathrm{g,m}}} &P_x\left(M_{\mathrm{s},i}\right) P_y\left(M_{\mathrm{s},j}\right)w_{\mathrm{p,ss}}\left(r_{\mathrm{p}}|M_{\mathrm{s},i},M_{\mathrm{s},j}\right), \label{equ:wp_xy}
\end{aligned}
\end{eqnarray}

where $i,j$ denote different halo (subhalo) bins. The probabilities that the central and satellite galaxies in the $\M$-selected subsample $x$ are hosted by the halo with $M_{\mathrm{h},i}$ and subhalo with $M_{\mathrm{s},i}$ are expressed as
\begin{eqnarray}
\begin{aligned}
P_{x}\left(M_{\mathrm{h},i}\right) &= P_{\mathrm{cen},x}\left(M^{\mathrm{min}}_{\ast,x}<\M<M^{\mathrm{max}}_{\ast,x}|M_{\mathrm{h},i}\right) \\
 &= \int_{M^{\mathrm{min}}_{\ast,x}}^{M^{\mathrm{max}}_{\ast,x}} p\left(\M|M_{\mathrm{h},i}\right) \mathrm{d} \M \label{equ:P_M(M_h)}
\end{aligned}
 \end{eqnarray}
and
\begin{eqnarray}
\begin{aligned}
P_{x}\left(M_{\mathrm{s},i}\right) &= P_{\mathrm{sat},x}\left(M^{\mathrm{min}}_{\ast,x}<\M<M^{\mathrm{max}}_{\ast,x}|M_{\mathrm{s},i}\right) \\
&= \int_{M^{\mathrm{min}}_{\ast,x}}^{M^{\mathrm{max}}_{\ast,x}} p\left(\M|M_{\mathrm{s},i}\right) \mathrm{d} \M, \label{equ:P_M(M_s)}
\end{aligned}
\end{eqnarray}
where $M^{\mathrm{min}}_{\ast,x}$ ($M^{\mathrm{max}}_{\ast,x}$) is the lower (upper) boundary of the $\M$-selected subsample $x$ and $p(\M|M)$ is the conditional PDF of the stellar mass defined in Equation \ref{equ:p(M_star|M)}. The modeled number density of the subsample $x$ can be calculated by
\begin{eqnarray}
n^{x}_{\mathrm{g,m}} = \sum_{i}\left[\bar{n}_{\mathrm{h},i}P_{x}\left(M_{\mathrm{h},i}\right) + \bar{n}_{\mathrm{s},i}P_{x}\left(M_{\mathrm{s},i}\right) \right]. 
\end{eqnarray}

\subsection{Fitting procedure} \label{Fitting procedure}
In the observation, we measure four auto correlation functions ($\boldsymbol{w}^{M0M0}_{\mathrm{p}}$, $\boldsymbol{w}^{M1M1}_{\mathrm{p}}$, $\boldsymbol{w}^{M2M2}_{\mathrm{p}}$ and $\boldsymbol{w}^{M3M3}_{\mathrm{p}}$), six cross correlation functions ($\boldsymbol{w}^{M0M1}_{\mathrm{p}}$, $\boldsymbol{w}^{M0M2}_{\mathrm{p}}$, $\boldsymbol{w}^{M0M3}_{\mathrm{p}}$, $\boldsymbol{w}^{M1M2}_{\mathrm{p}}$, $\boldsymbol{w}^{M1M3}_{\mathrm{p}}$, $\boldsymbol{w}^{M2M3}_{\mathrm{p}}$) and four galaxy number densities ($n^{M0}_{\mathrm{g}}$, $n^{M1}_{\mathrm{g}}$, $n^{M2}_{\mathrm{g}}$ and $n^{M3}_{\mathrm{g}}$) for the stellar mass-selected subsamples. For the correlation function between subsamples $Mi$ and $Mj$, we can define its $\chi^2$ as 
\begin{eqnarray}
\begin{aligned}
&\chi^2_{MiMj} \\
&=\sum_{k=1}^{N_{\mathrm{r_p}}}\sum_{l=1}^{N_{\mathrm{r_p}}}\left(w_{\mathrm{p},k}-w_{\mathrm{p,m},k}\right)C^{-1}_{kl}\left(w_{\mathrm{p},l}-w_{\mathrm{p,m},l}\right), \label{equ:chi2_MiMj}
\end{aligned}
\end{eqnarray}
where $\boldsymbol{w}_{\mathrm{p}}$, $\boldsymbol{w}_{\mathrm{p,m}}$, and $\boldsymbol{C}$ denote the observed correlation function $\boldsymbol{w}^{MiMj}_{\mathrm{p}}$, the model prediction $\boldsymbol{w}^{MiMj}_{\mathrm{p,m}}$, and the covariance matrix $\boldsymbol{C}^{MiMj}$, respectively. Here the inverse of covariance matrix $\boldsymbol{C}^{-1}$ is multiplied by a bias-correction factor $(N_{\mathrm{jack}}-N_{\mathrm{r_p}}-2)/(N_{\mathrm{jack}}-1)$ \citep{2007A&A...464..399H}, where $N_{\mathrm{jack}}=24$ and $N_{\mathrm{r_p}}=12$ are the number of jackknife subsamples and $r_{\mathrm{p}}$ bins respectively. Then the total $\chi^2$ is written as
\begin{eqnarray}
\begin{aligned}
\chi^2=\sum_{i=0}^{3}\sum_{j=i}^{3}\chi^2_{MiMj} + \sum_{i=0}^{3}\frac{\left(n^{Mi}_{\mathrm{g}}-n^{Mi}_{\mathrm{g,m}}\right)^{2}}{\sigma^{2}_{Mi}}, \label{equ:chi2}
\end{aligned}
\end{eqnarray}
 where $n^{Mi}_{\mathrm{g,m}}$ is the modeled number density of the $i$-th subsample and $\sigma_{Mi}$ is the field-to-field variation in different jackknife fields. Particularly, since the red satellite galaxies in the $M0$ subsample may be slightly incomplete at $z>0.7$, the one-halo term of the cross correlation between $M0$ and $M3$, which mainly contains the massive central galaxies, is more likely to be suppressed if the red satellite galaxies in $M0$ are missing. Conservatively, we remove $\boldsymbol{w}^{M0M3}_{\mathrm{p}}$ (corresponding to $i=0, j=3$ in Equation \ref{equ:chi2}) in our fitting. In addition, considering the current limited data size, we ignore the covariance between different subsamples and use a total of nine covariance matrices each with $12\times12$ elements in Equation \ref{equ:chi2_MiMj}. The degree of freedom is therefore ${\rm dof}=12\times9 + 4 -5 = 107$.  In Bayesian theory, the posterior distribution is proportional to the likelihood function times the prior of the parameters. We set wide priors for the five parameters: $10<\log M_0<13$, $0.1<\alpha<0.5$, $1<\beta<5$, $9<\log k<12$ and $0<\sigma<1$. An Markov Chain Monte Carlo (MCMC) analysis is performed with {\tt\string emcee} \citep{2013PASP..125..306F}.  The posterior distributions of the model parameters are shown in Figure \ref{fig:fit_corner_cross_mcmc_SameCS_9SM_table_with_sigmaz_wp_ZMAX_SM_omegam0268}. Overall, all the five parameters are well determined.

We present the best-fitting $\boldsymbol{w}_{\mathrm{p}}$ as well as the SHMR as solid lines in Figure \ref{fig:bestfitting_cross_mcmc_SameCS_9SM_table_with_sigmaz_wp_ZMAX_SM_omegam0268}. The reduced $\chi^2$ is equal to 1.12, indicating a good overall fit. Compared to the SHMR model derived by \cite{2010MNRAS.402.1796W} at $z\sim 0.8$ using VVDS observation \citep{2007A&A...474..443P}, $\log M_0$ is slightly larger and $\alpha$ is slightly smaller in our model ($\log M_0=11.64$ and $\alpha=0.29$ in their unified model). This is partly because there is a degeneracy of $\log M_0$ and $\alpha$, which 
 is also evident in the $\log M_0$-$\alpha$ contour in Figure \ref{fig:fit_corner_cross_mcmc_SameCS_9SM_table_with_sigmaz_wp_ZMAX_SM_omegam0268}. Nevertheless, with the larger galaxy sample of VIPERS, we have imposed tighter constraints on the SHMR model at $z\sim 0.6$. 

\begin{figure}
	\centering
	\includegraphics[scale=0.75]{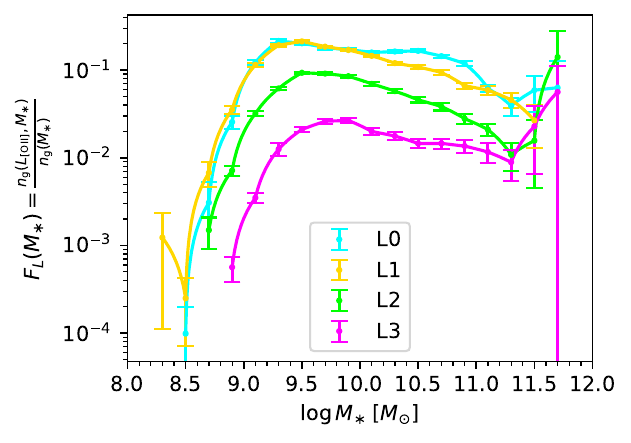}
	\caption{The fractions of the ELGs in the $L_\oiirm$-selected subsamples, in terms of the whole galaxy population,  as a function of stellar mass. The measured $F_L\left(\M\right)$ as well as its Poisson error for different subsamples are denoted as the data points with different colors. The linear interpolation of the observed $F_L\left(\M\right)$-$\log \M$ relationships are shown as solid lines with corresponding colors.   
		\label{fig:SM_LOIIweighted_by_Vmax_to_SMsmf_McLeod_relation_bin}}
\end{figure}

\section{Construct the halo occupation of ELGs} \label{sec:populate ELGs}
In this section, we aim to propose an efficient way to construct the ELG-halo connection. We investigate how to populate the halos with ELGs in the simulation with the measured ELG-stellar mass relation and SHMR. We test our method in both real-space and redshift-space. We also propose a model for HOD modeling of ELGs.
\subsection{ELG-stellar mass relation in the observation} \label{sec:The ELG-stellar mass relation in observation}
We first measure the fraction of ELGs in the whole population of galaxies as a function of stellar mass in the observation. For each $L_\oiirm$-selected subsample, we divide the galaxies into twenty $\log \M$ bins ranging from $\log \M = 8$ to $12\,M_{\odot}$ with a bin width $\Delta \log \M = 0.2$, and compute the weighted number density $n_{\mathrm{g}}\left(L_\oiirm,\M\right)$ in each bin using Equation \ref{equ:number density}. Then the fraction of each $L_\oiirm$-selected subsample at a given stellar mass is defined as 
\begin{eqnarray}
F_L\left(\M\right) = \frac{n_{\mathrm{g}}\left(L_\oiirm,\M\right)}{n_{\mathrm{g}}\left( \M \right)}. \label{equ:fraction}
\end{eqnarray}

where the number density $n_{\mathrm{g}}\left( \M \right)$ of {\it all} galaxies for a stellar mass bin can be estimated by integrating the galaxy stellar mass function (SMF) $\Phi\left(\M\right)$:
\begin{eqnarray}
n_{\mathrm{g}}\left( \M \right) = \int_{\log \M - \Delta \log \M/2}^{\log \M + \Delta \log \M/2} \Phi\left(\M\right) \mathrm{d}\log \M.
\end{eqnarray}

Here we adopt the SMF measured by \cite{2021MNRAS.503.4413M} in the redshift range $0.25<z<0.75$. \cite{2021MNRAS.503.4413M} has combined the data from the Hubble Space Telescope (HST) CANDELS fields \citep{2011ApJS..197...35G,2011ApJS..197...36K} and other ground-based surveys with deep photometric measurements, and provided the best-fitting parameters of the double Schechter function \citep{1976ApJ...203..297S}. 

In Figure \ref{fig:SM_LOIIweighted_by_Vmax_to_SMsmf_McLeod_relation_bin}, we show the fraction $F_L\left(\M\right)$ for different $L_\oiirm$-selected subsamples. The error bars represent the Poisson errors of the weighted number counts. Firstly, we note that the shapes of $F_L\left(\M\right)$ for the four subsamples are similar, while the locations of the peaks of $F_L\left(\M\right)$ shift slightly from $10^{9.3}$ to $10^{9.7} \,M_{\odot}$ with \oii luminosity increasing. Furthermore, at the low-mass end, the $i$-band magnitude limit may have led to a rapid decrease of the number of galaxies, thus causing the rapid drop of $F_L\left(\M\right)$. The gradual decrease of $F_L\left(\M\right)$ at the high-mass is expected, because more massive galaxies are more likely to stop their star formation and become quiescent. Moreover, the $F_L\left(\M\right)$ of both $L2$ and $L3$ show an upturn at $\log \M >11.3 \,M_{\odot}$. This feature might imply that the galaxies at the high-mass end are likely to host AGN, whose violent activities are sufficient enough to generate strong \oii emissions \citep[e.g.,][]{2011ApJ...737L..38K}. Nevertheless, the AGN contamination cannot significantly affect our results because the number of the massive galaxies with $\log \M >11.3 \,M_{\odot}$ is very small (only 12 galaxies in $L2$ and 10 galaxies in $L3$). Instead of using a parameterized model, we linearly interpolate the $F_L\left(\M\right)$-$\log \M$ relationships to preserve the observed information. Moreover, it is worth mentioning that \cite{2019ApJ...871..147G} points out that the completeness of the ELG sample in eBOSS varies from 1\% to 10\%  at different stellar masses, which is comparable to the $F_L\left(\M\right)$ of the $L2$ or $L3$ subsample. This is due to the $g$-band magnitude limit of eBOSS \citep{2017MNRAS.471.3955R}, which causes the majority of the selected ELGs to be luminous ones.

\begin{figure*}
	\centering
	\includegraphics[scale=0.6]{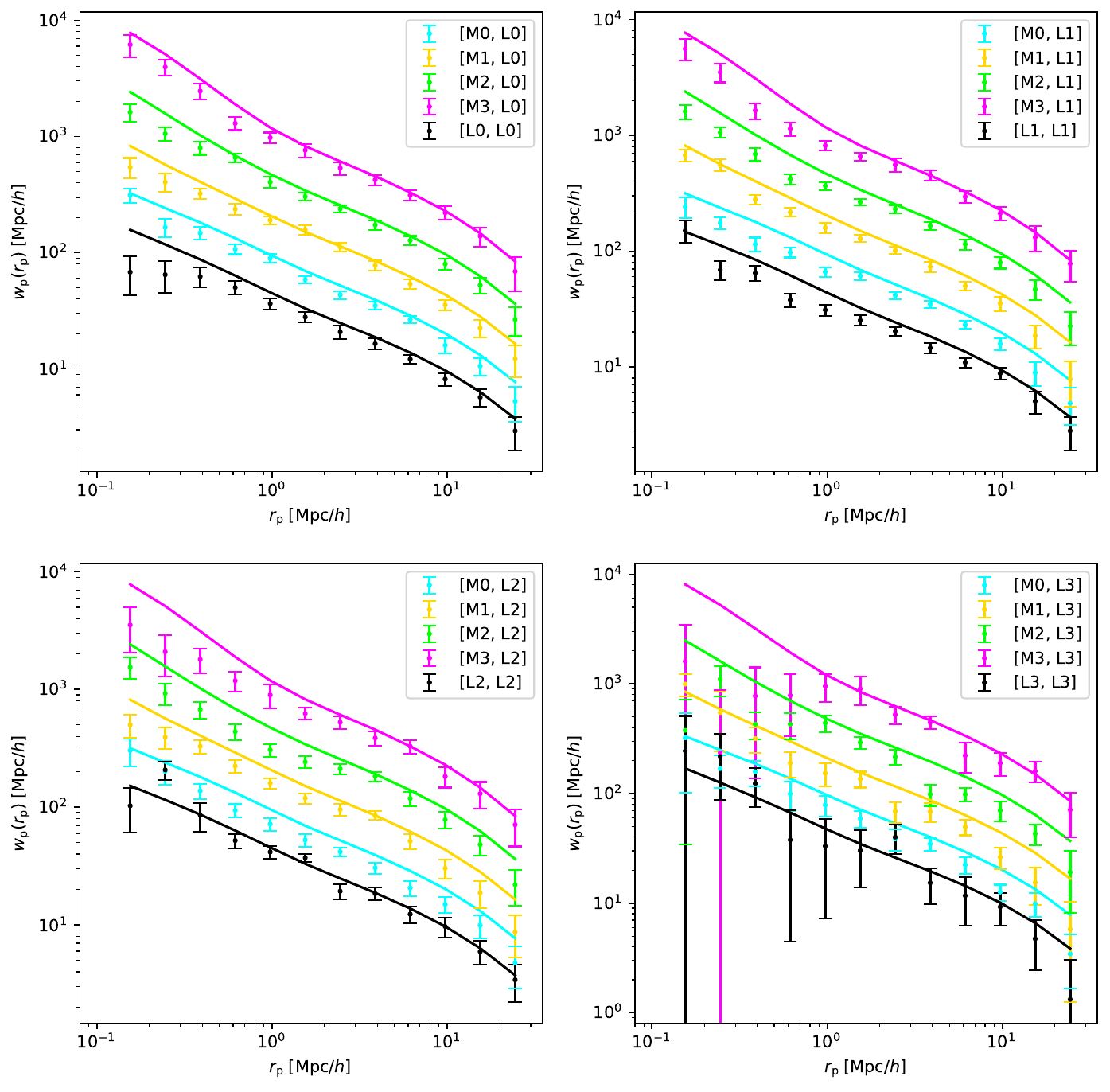}
	
	\caption{The observed projected cross (auto) correlation functions for the $L_\oiirm$-selected subsamples, compared with the AM model predictions by randomly populating ELGs according to the SHMR and the fraction $F_L\left(\M\right)$. The results of the four subsamples $L0$, $L1$, $L2$ and $L3$ are shown in four panels respectively. The data points with error bars are measured from VIPERS. The model predictions are plotted as solid lines. Except for the auto correlations (black), all the other cross correlations have been multiplied by $2^n$ where $n$ changes with color ($n = $ 1 (cyan), 2 (yellow), 3 (lime) and 4 (magenta)) to give a clear illustration.   
		\label{fig:nofsat_sim_cross_randomSM_SameCS_9SM_table_with_sigmaz_wp_ZMAX_LOII_omegam0268.pdf}}
\end{figure*}

\begin{figure*}
	\centering
	\includegraphics[scale=0.6]{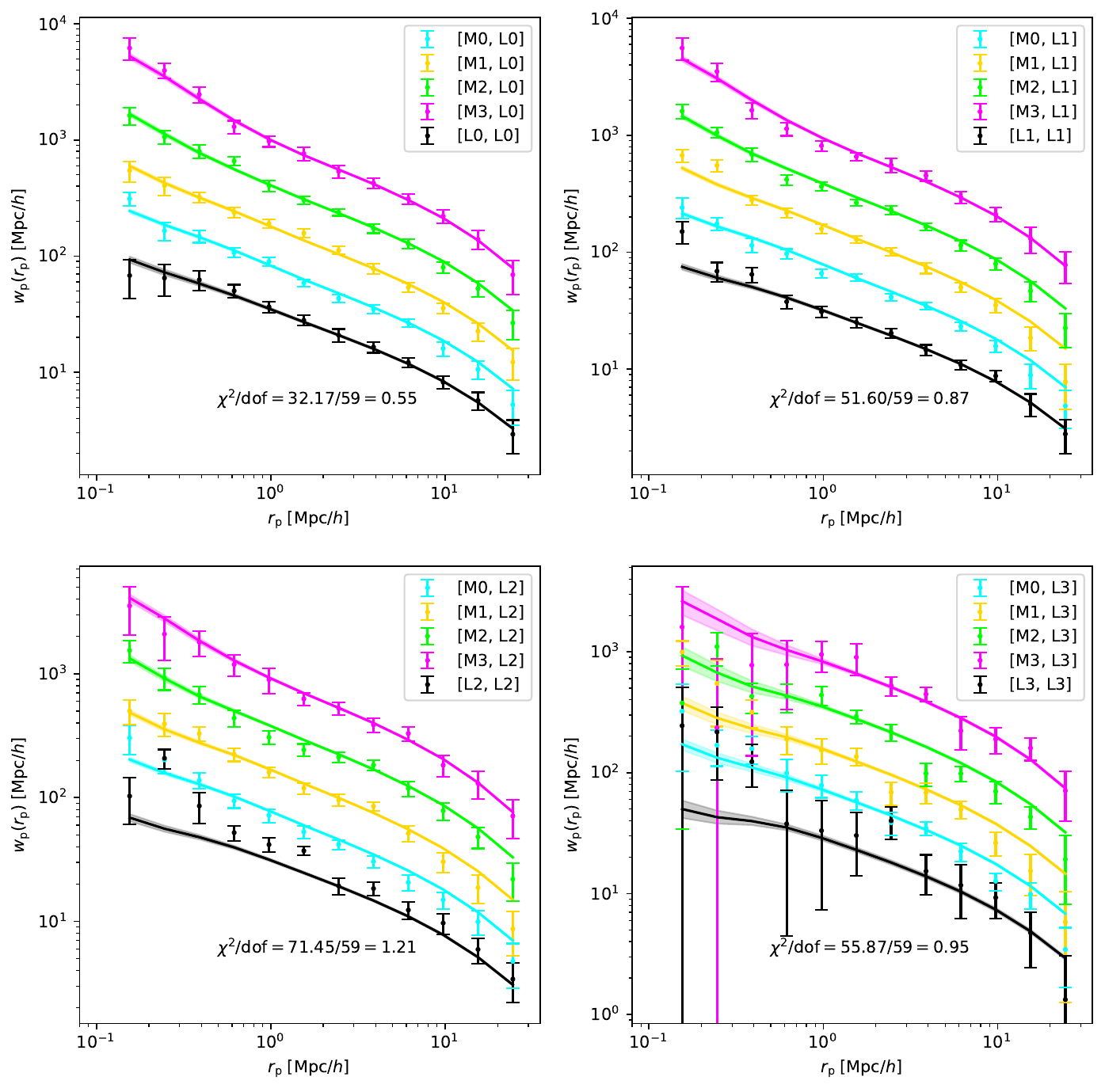}
	
	\caption{Similar to Figure \ref{fig:nofsat_sim_cross_randomSM_SameCS_9SM_table_with_sigmaz_wp_ZMAX_LOII_omegam0268.pdf}, but the lines are the fitting results of our model with an adjustable satellite fraction $f_{\mathrm{sat}}$. The best-fitting models as well as their $1\sigma$ error are plotted as solid lines with shadow areas. The reduced $\chi^2$ is also marked in each panel.  
		\label{fig:bestfitting_sim_cross_randomSM_SameCS_9SM_table_with_sigmaz_wp_ZMAX_LOII_omegam0268}}
\end{figure*}

\begin{figure}
	\centering
	\includegraphics[scale=0.85]{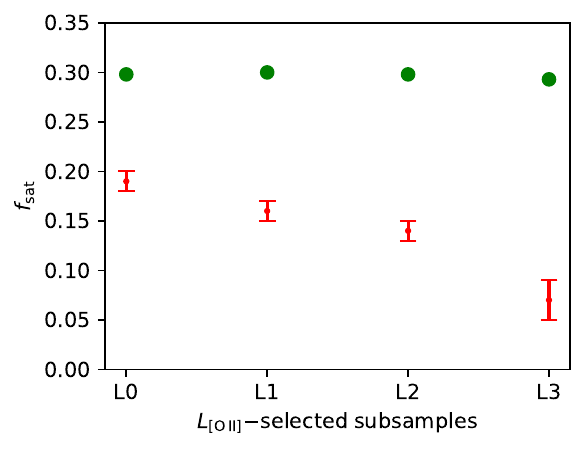}
	\caption{Our best-fitting satellite fractions $f_{\mathrm{sat}}$ for the four $L_\oiirm$-selected subsamples (red points with error bars). For comparison, the green circles represent the $f_{\mathrm{sat}}$ if each $L_\oiirm$-selected subsample is selected randomly based on the SHMR and $F_L\left(\M\right)$.		\label{fig:f_sat}}
	
\end{figure}
\subsection{Populating halos with ELGs} \label{sec:Assigning ELGs with abundance matching}
Combining the ELG-stellar mass relation measured above and the SHMR derived in Section \ref{sec:SHMR}, we can connect the ELGs with halos in the simulation. To model the auto correlation functions of the $L_\oiirm$-selected subsamples and their cross correlation functions with the $\M$-selected subsamples, we also adopt Equation \ref{equ:wp_xy} to make the calculation more efficient, just changing the $P_{x\left(y\right)}\left(M_{\mathrm{h},i}\right)$ and $P_{x\left(y\right)}\left(M_{\mathrm{s},i}\right)$ in the Equation. For a $\M$-selected subsample $x$, the $P_{x}\left(M_{\mathrm{h},i}\right)$ and $P_{x}\left(M_{\mathrm{s},i}\right)$ can also be calculated by Equation \ref{equ:P_M(M_h)} and \ref{equ:P_M(M_s)}, respectively. And for a $L_\oiirm$-selected subsample $x$, we can define its $P_{x}\left(M_{\mathrm{h},i}\right)$ and $P_{x}\left(M_{\mathrm{s},i}\right)$ as

\begin{eqnarray}
\begin{aligned}
&P_{x}\left(M_{\mathrm{h},i}\right)\\
 &= P_{\mathrm{cen},x}\left(L^{\mathrm{min}}_{\oiirm,x}<L_\oiirm<L^{\mathrm{max}}_{\oiirm,x}|M_{\mathrm{h},i}\right) \\
&= \int_{-\infty}^{+\infty} F_{Lx}\left(\M\right)p\left(\M|M_{\mathrm{h},i}\right)\mathrm{d}\M \label{equ:P_L(M_h)}
\end{aligned}
\end{eqnarray}
and
\begin{eqnarray}
\begin{aligned}
&P_{x}\left(M_{\mathrm{s},i}\right)\\
&= P_{\mathrm{sat},x}\left(L^{\mathrm{min}}_{\oiirm,x}<L_\oiirm<L^{\mathrm{max}}_{\oiirm,x}|M_{\mathrm{s},i}\right) \\
&= \int_{-\infty}^{+\infty} F_{Lx}\left(\M\right)p\left(\M|M_{\mathrm{s},i}\right)\mathrm{d}\M, \label{equ:P_L(M_s)}
\end{aligned}
\end{eqnarray}
where the $F_{Lx}\left(\M\right)$ is the fraction of subsample $x$ at given $\M$ and $p\left(\M|M_{\mathrm{h},i}\right)$  ($p\left(\M|M_{\mathrm{s},i}\right)$) is fixed to the best-fitting SHMR derived in Section \ref{sec:SHMR}. In the above equations, the sample of ELGs is equivalent to a random selection of the fraction $F_{Lx}\left(\M\right)$ of galaxies from the whole population. In this manner, we calculate the modeled projected cross (auto) correlation functions $w_{\mathrm{p}}$ for each $L_\oiirm$-selected subsample and present them as solid lines in Figure \ref{fig:nofsat_sim_cross_randomSM_SameCS_9SM_table_with_sigmaz_wp_ZMAX_LOII_omegam0268.pdf}, where the four panels represent the four $L_\oiirm$-selected subsamples. We note that our model overestimates the overall clustering signal, especially at small scales. This may be caused by the assumption that the satellite fraction in each subsample is the same as that of the normal galaxies. However, in the real Universe, satellite galaxies are expected form earlier than central ones, so the probability that they are currently star-forming ELGs is relatively lower. Furthermore, the modeled satellite fractions $f_{\mathrm{sat}}$ are displayed as green circles in Figure \ref{fig:f_sat}. The value of satellite fraction is close to 0.3, which is obviously higher than that found in current observational studies \citep[e.g.,][]{2019ApJ...871..147G, 2021PASJ...73.1186O}, in which $f_{\mathrm{sat}}<0.2$.   

Motivated by these considerations, we introduce a free parameter $f_{\mathrm{sat}}$ to modulate the satellite fraction in our model. The Equation \ref{equ:P_L(M_h)} and \ref{equ:P_L(M_s)} are re-written as
\begin{eqnarray}
\begin{aligned}
&P_{x}\left(M_{\mathrm{h},i}\right)\\
&= P_{\mathrm{cen},x}\left(L^{\mathrm{min}}_{\oiirm,x}<L_\oiirm<L^{\mathrm{max}}_{\oiirm,x}|M_{\mathrm{h},i}\right) \\
&= \left(1-f_{\mathrm{sat}}\right)\frac{\bar{n}_{\mathrm{h},i}+\bar{n}_{\mathrm{s},i}}{\bar{n}_{\mathrm{h},i}}\times\int_{-\infty}^{+\infty} F_{Lx}\left(\M\right)p\left(\M|M_{\mathrm{h},i}\right)\mathrm{d}\M \label{equ:P_L(M_h)_fsat}
\end{aligned}
\end{eqnarray}
and
\begin{eqnarray}
\begin{aligned}
&P_{x}\left(M_{\mathrm{s},i}\right)\\
&= P_{\mathrm{sat},x}\left(L^{\mathrm{min}}_{\oiirm,x}<L_\oiirm<L^{\mathrm{max}}_{\oiirm,x}|M_{\mathrm{s},i}\right) \\
&= f_{\mathrm{sat}}\frac{\bar{n}_{\mathrm{h},i}+\bar{n}_{\mathrm{s},i}}{\bar{n}_{\mathrm{s},i}}\times\int_{-\infty}^{+\infty} F_{Lx}\left(\M\right)p\left(\M|M_{\mathrm{s},i}\right)\mathrm{d}\M, \label{equ:P_L(M_s)_fsat}
\end{aligned}
\end{eqnarray}

where the $f_{\mathrm{sat}}$ is the satellite fraction. After the SHMR $p\left(\M|M\right)$ is fixed with the best-fitting parameters shown in Figure \ref{fig:fit_corner_cross_mcmc_SameCS_9SM_table_with_sigmaz_wp_ZMAX_SM_omegam0268}, the ELG-halo connection can be determined completely by the only one free parameter $f_{\mathrm{sat}}$.

Next we constrain the parameter $f_{\mathrm{sat}}$ by fitting our model with the observed cross (auto) correlation functions. Similar to Equation \ref{equ:chi2}, $\chi^2$ for the $i$-th $L_\oiirm$-selected subsample $Li$ can be written as
\begin{eqnarray}
\begin{aligned}
\chi_{Li}^2  = \chi_{LiLi}^2 + \sum_{j=0}^{3}\chi_{LiMj}^2, \label{equ:chi2_L}
\end{aligned}
\end{eqnarray}
where we use one auto correlation function $\boldsymbol{w}^{LiLi}_{\mathrm{p}}$ and four cross correlation functions $\boldsymbol{w}^{LiMj}_{\mathrm{p}}$ to infer the model parameter $f_{\mathrm{sat}}$. The corresponding $\chi_{LiLi}^2$ and $\chi_{LiMj}^2$ can also be computed in analog to Equation \ref{equ:chi2_MiMj}. As a result, the dof in our fitting is $\mathrm{dof} = 12\times5-1=59$. 

We show the best-fitting $f_{\mathrm{sat}}$ as well as the $1\sigma$ dispersion of their posterior distributions in Figure \ref{fig:f_sat}. It demonstrates that the best-fitting $f_{\mathrm{sat}}$ decreases as the $L_\oiirm$ increases, indicating that the \oii lines are primarily generated by central galaxies rather than old satellites with little star formation. The value of $f_{\mathrm{sat}}$ in our model is also broadly consistent with other observational results \citep[e.g.,][]{2019ApJ...871..147G, 2021PASJ...73.1186O}.   

The best-fitting $w_{\mathrm{p}}$ as well as the $1\sigma$ uncertainties are displayed as the solid lines with shadow regions in Figure \ref{fig:bestfitting_sim_cross_randomSM_SameCS_9SM_table_with_sigmaz_wp_ZMAX_LOII_omegam0268}. The cross correlation functions of the four ELGs subsamples are well fitted. 
It suggests that the SHMR of normal galaxies can be used for ELGs with only the fraction of satellite galaxies reduced. This may indicate that the clustering of normal galaxies in the stellar mass range of ELGs does not depend on the star formation rate. As the ELGs are mostly in the stellar mass range $<10^{10}\,M_\odot$, we expect that normal central galaxies in this mass range at redshift $z\sim 0.7$ are dominantly star forming galaxies, which supports why we can use the same SHMR. The lower fraction of the satellites indicates that the red satellites should not be included in the ELG sample. Our results are also broadly consistent with the finding of \cite{2021MNRAS.502.3599H} that DESI-like ELGs have a small assembly bias based on IllustrisTNG simulations. In addition, we note that the observed auto correlations of $L2$ and $L3$ are slightly higher than our model predictions especially at small scales, although the errors are large. In the future, we will carefully investigate this issue using a much larger ELG sample from DESI.  

\begin{figure*}
	\centering
	\includegraphics[scale=0.5]{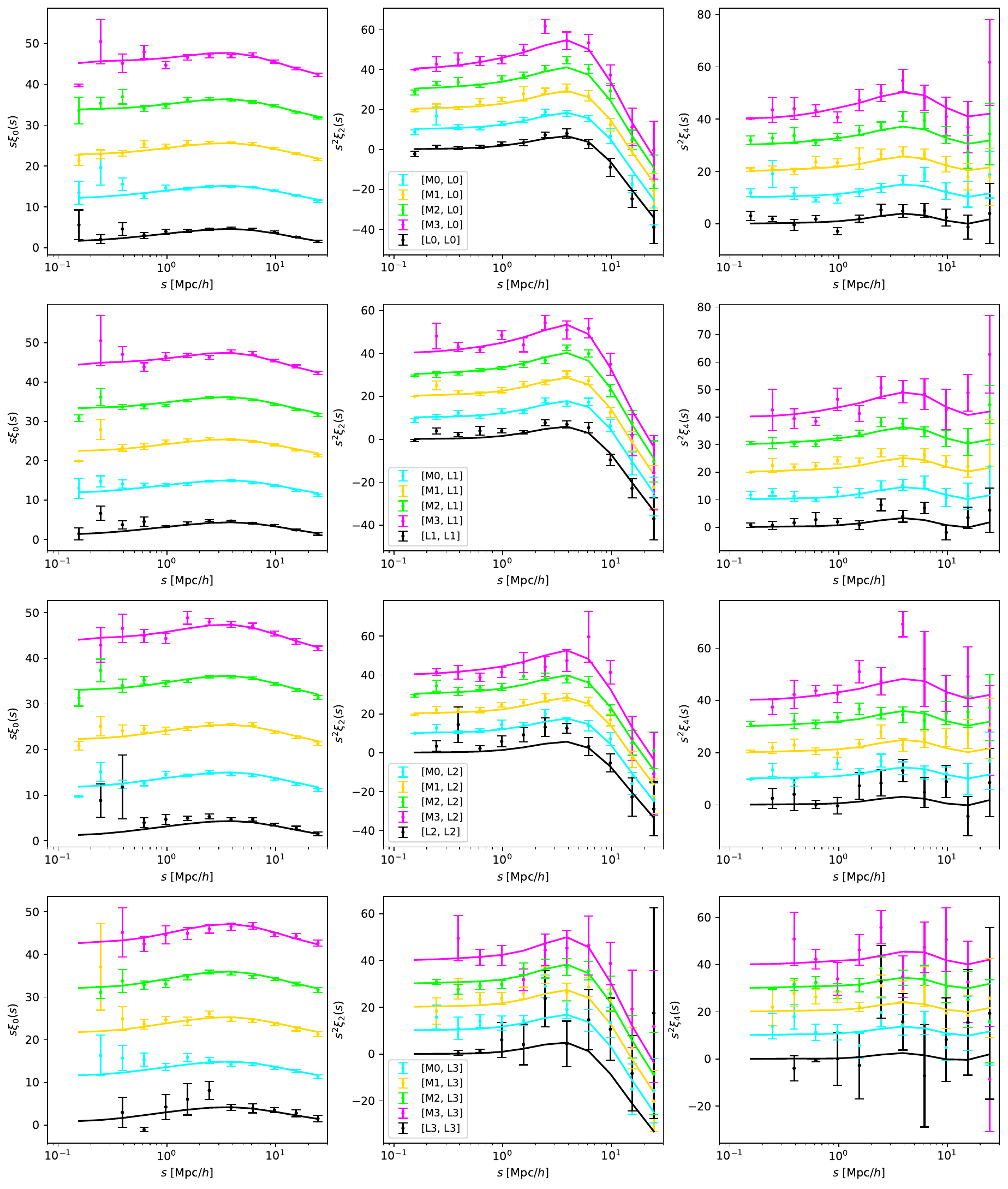}
	\caption{The cross (auto) correlation functions in redshift-space for the $L_\oiirm$-selected subsamples in both the observations and our models. The four rows from top to bottom represent the four subsamples $L0$, $L1$, $L2$ and $L3$. The observed monopole $s\xi_0\left(s\right)$, quadrupole $s^2\xi_2\left(s\right)$ and hexadecapole $s^2\xi_4\left(s\right)$ are displayed as data points with error bars in the three columns from left to right, respectively. Except for the black points, all the other data points have been shifted by $10\times n$ where $n$ changes with color ($n = $ 1 (cyan), 2 (yellow), 3 (lime) and 4 (magenta)) to give a clear display. The solid lines represent our model predictions ({\it not fittings}). 
		\label{fig:bestfitting_sim_cross_randomSM_SameCS_9SM_table_with_sigmaz_xi024shift_ZMAX_LOII_omegam0268}}
\end{figure*}

\begin{figure*}
	\centering
	\includegraphics[scale=0.5]{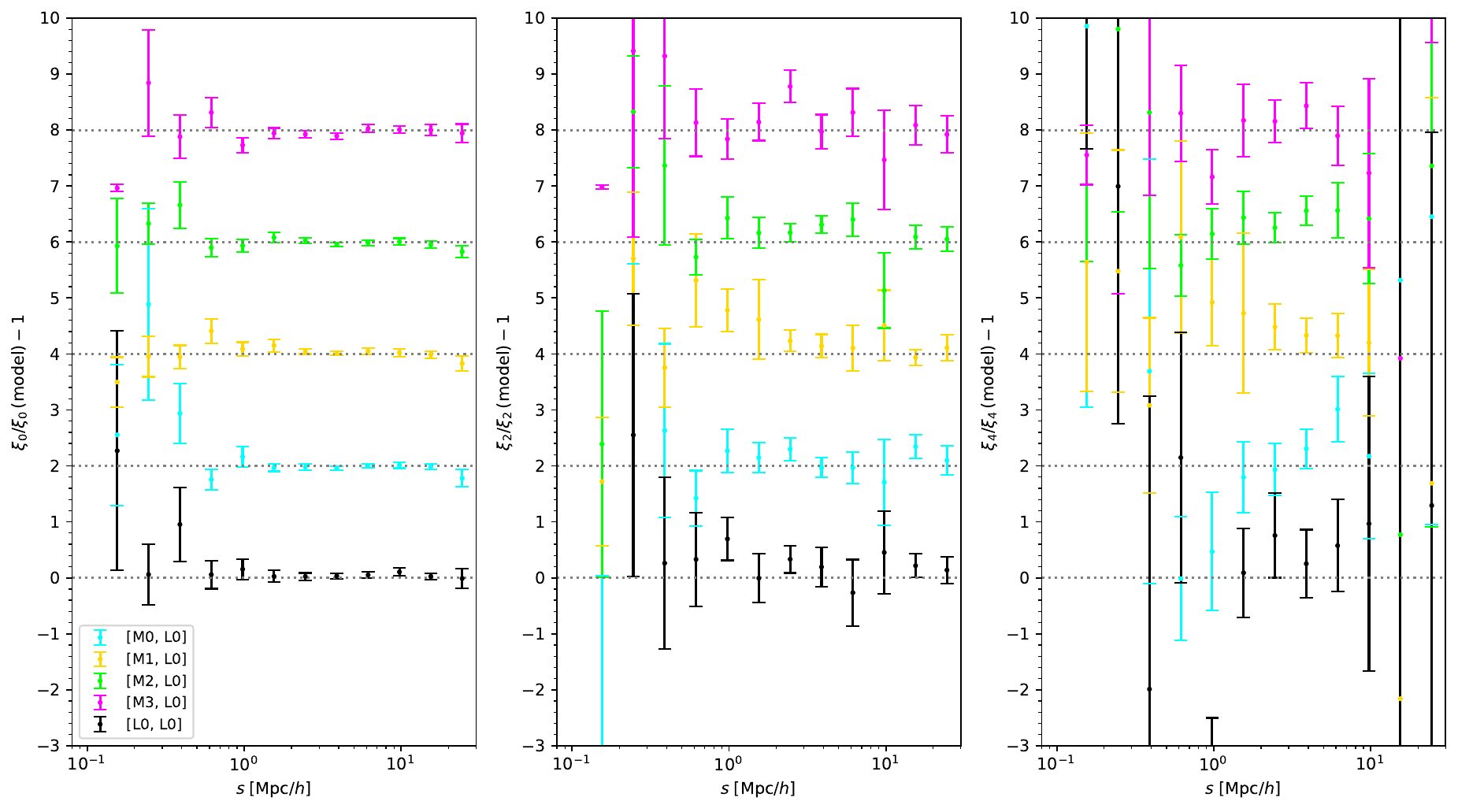}
	\caption{Similar to the top row of Figure \ref{fig:bestfitting_sim_cross_randomSM_SameCS_9SM_table_with_sigmaz_xi024shift_ZMAX_LOII_omegam0268}, but the points with error bars represent the ratios of the multiple moments between the observations and the model predictions for the subsample $L0$. Except for the black points, all the other data points have been shifted by $2\times n$ where $n$ changes with color ($n = $ 1 (cyan), 2 (yellow), 3 (lime) and 4 (magenta)) to give a clear display.   
		\label{fig:bestfitting_sim_cross_randomSM_SameCS_9SM_table_with_sigmaz_ratio_L0_ZMAX_LOII_omegam0268}}
\end{figure*}

\subsection{Predicting the correlation functions of ELGs in redshift-space} \label{sec:Predicting the correlation functions of ELGs in redshift-space}
We further check the performance of our model predictions for the clustering in redshift-space. By replacing the $w_{\mathrm{p}}\left(r_{\mathrm{p}}\right)$ in Equation \ref{equ:wp_xy} with $\xi_0\left(s\right)$, $\xi_2\left(s\right)$ and $\xi_4\left(s\right)$, we can calculate these predicted multipole moments, which are presented in Figure \ref{fig:bestfitting_sim_cross_randomSM_SameCS_9SM_table_with_sigmaz_xi024shift_ZMAX_LOII_omegam0268} as solid curves. Although we have only fitted the observed real-space $w_{\mathrm{p}}\left(r_{\mathrm{p}}\right)$, the multipole moments in redshift-space predicted by our best-fitting model are also in good agreement with the observations. The ratios of the multiple moments between the observations and the model for the subsample $L0$ (top row in Figure \ref{fig:bestfitting_sim_cross_randomSM_SameCS_9SM_table_with_sigmaz_xi024shift_ZMAX_LOII_omegam0268}) are shown in Figure \ref{fig:bestfitting_sim_cross_randomSM_SameCS_9SM_table_with_sigmaz_ratio_L0_ZMAX_LOII_omegam0268}. We can notice that the relative difference between the observed $\xi_0\left(s\right)$ and our model is always about 1$\sigma$, and less than $\sim 10\%$ for those well-measured data points. For $\xi_2\left(s\right)$ and $\xi_4\left(s\right)$, although there are larger uncertainties in the measurements, the overall relative difference is still within about $1\sigma$ confidence interval. We omit figures for the other luminosity subsamples, since their ratios have behaviors similar to what shown from the subsample $L0$.

In general, our model can well reproduce the cross (auto) correlation functions in both real-space and redshift-space for the ELGs. This method can be easily applied to generate ELG mock catalogs for ongoing spectroscopic surveys such as DESI and PFS.

\begin{figure}
	\centering
	\includegraphics[scale=0.7]{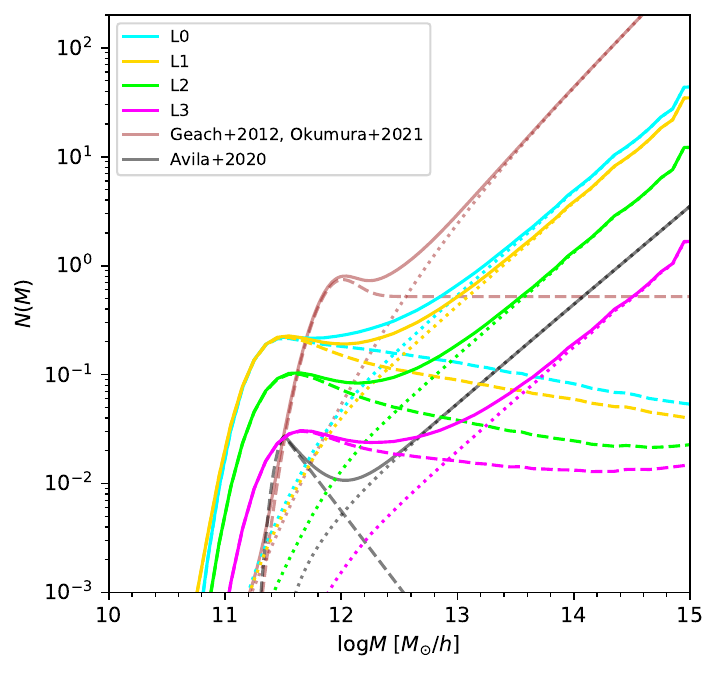}
	\caption{The comparison of different ELG HOD models. The predicted HODs of our four \oii luminosity-selected subsamples are presented as cyan, yellow, lime and magenta lines. In addition, we present the Geach HOD model \citep{2012MNRAS.426..679G} with the best-fitting parameters provided by \cite{2021PASJ...73.1186O} as brown lines. The HOD model proposed by \cite{2020MNRAS.499.5486A} is also shown as black lines. The solid, dashed and dotted lines denote the total, central and satellite occupation numbers, respectively. 
		\label{fig:bestfittingHOD_logM50_compare_sim_cross_randomSM_SameCS_9SM_table_with_sigmaz_wp_ZMAX_LOII_omegam0268}}
\end{figure}

\begin{figure*}
	\centering
	\includegraphics[scale=0.6]{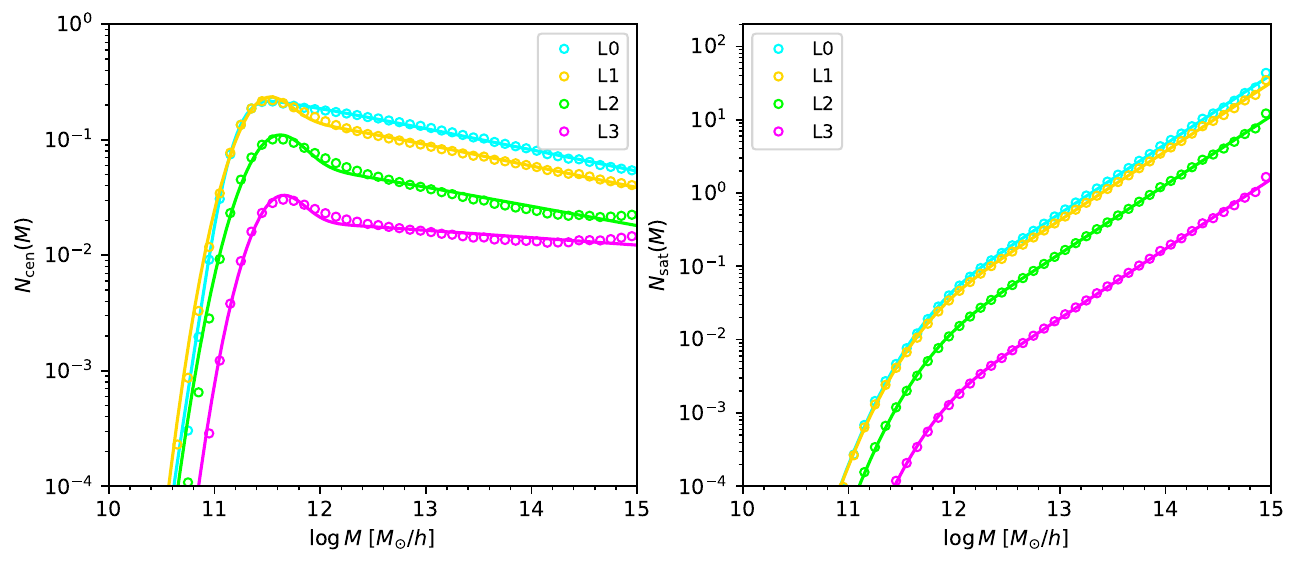}
	\caption{Comparison of the modified Geach HOD form (Equation \ref{equ:newform}) with the occupations derived from our mock catalog. The hollow circles denote our derived HODs (same as those shown in Figure \ref{fig:bestfittingHOD_logM50_compare_sim_cross_randomSM_SameCS_9SM_table_with_sigmaz_wp_ZMAX_LOII_omegam0268}) of our mock catalogs for the four $L_\oiirm$ -selected subsamples. The best-fitting results are shown as solid lines. The left and right panels correspond to $N_{\mathrm{cen}}$ and $N_{\mathrm{sat}}$ respectively. 
		\label{fig:plot_modified_HOD_9p_compare}}
\end{figure*}

\subsection{Implications for HOD modeling} \label{sec:Comparison of HOD models}
In order to compare with the traditional HOD models, we work out the HOD based on our model,
\begin{eqnarray}
\begin{aligned}
&N_{\mathrm{cen},x} \left(M_{\mathrm{h}}\right) = P_{x}\left(M_{\mathrm{h}}\right) \\
&N_{\mathrm{sat},x} \left(M_{\mathrm{h}}\right) = \int  P_{x}\left(M_{\mathrm{s}}\right) n_{\mathrm{sub}}\left(M_{\mathrm{s}}|M_{\mathrm{h}}\right) \mathrm{d}M_{\mathrm{s}} \\
&N_{x} \left(M_{\mathrm{h}}\right) = N_{\mathrm{cen},x} \left(M_{\mathrm{h}}\right)+N_{\mathrm{sat},x} \left(M_{\mathrm{h}}\right),
\end{aligned}
\end{eqnarray}

where $N_{\mathrm{cen},x} \left(M_{\mathrm{h}}\right)$, $N_{\mathrm{sat},x} \left(M_{\mathrm{h}}\right)$ and $N_{x} \left(M_{\mathrm{h}}\right)$ are the central, satellite and total occupation numbers respectively in the ELG subsample $x$, and the probabilities $P_{x}\left(M_{\mathrm{h}}\right)$ and $P_{x}\left(M_{\mathrm{s}}\right)$ are calculated using Equation \ref{equ:P_L(M_h)_fsat} and \ref{equ:P_L(M_s)_fsat} but with a bin width $\Delta \log M = 0.1$. Here $n_{\mathrm{sub}}\left(M_{\mathrm{s}}|M_{\mathrm{h}}\right)$ measured from our simulation is the mean subhalo mass function at the given $M_{\mathrm{h}}$ bin. The HODs of the four \oii luminosity-selected subsamples are shown in Figure \ref{fig:bestfittingHOD_logM50_compare_sim_cross_randomSM_SameCS_9SM_table_with_sigmaz_wp_ZMAX_LOII_omegam0268} as solid lines, and the decomposed central and satellite occupation numbers are also displayed as dashed and dotted lines, respectively. Then we compare our model predictions with two recent HOD models of \oii ELGs. It should be noted that since the HOD depends on the target selections of the ELG samples, we can only qualitatively compare the shapes of these HODs instead of their precise values. 

The first is the Geach HOD model \citep{2012MNRAS.426..679G}. \cite{2021PASJ...73.1186O} has constrained the model parameters based on the HSC NB observations	of \oii emitters at $z=1.19$ and $z=1.47$, and found that the model can well fit the angular correlation functions of the \oii emitters. Considering that the parameters are better constrained at $=1.47$, we adopt their model parameters at this redshift based on the posterior PDF (see their Table 3) and display the HOD as brown lines in Figure \ref{fig:bestfittingHOD_logM50_compare_sim_cross_randomSM_SameCS_9SM_table_with_sigmaz_wp_ZMAX_LOII_omegam0268}. The shape of their $N_{\mathrm{cen}}$ at low-mass end is quite similar to ours. However, with the Geach HOD form,  $N_{\mathrm{cen}}$ in their model tends to be a constant at large halo mass, while our $N_{\mathrm{cen}}$ keeps decreasing. Although the HOD at massive end has a relatively small effect on galaxy clustering due to the rapid decline of the halo mass function, our results imply that a decreasing function $N_{\mathrm{cen}}=\propto M^{\beta_{\mathrm{c}}}$ ($\beta_{\mathrm{c}}\sim -0.2$) can better describe the massive end of ELG HOD. On the other hand, both the Geach model and ours present a similar power-law form for $N_{\mathrm{sat}}$. 

The other is the HOD model (the HOD-3 in their paper) proposed by \cite{2020MNRAS.499.5486A} for the eBOSS ELGs. This model combines a Gaussian function with a decaying power-law form to describe the central occupation. \cite{2020MNRAS.499.5486A} has constrained this model using the semi-analytical model (SAM) results \citep{2018MNRAS.474.4024G} as well as the eBOSS number density and bias (see their Table 2). We show this model as black curves in Figure \ref{fig:bestfittingHOD_logM50_compare_sim_cross_randomSM_SameCS_9SM_table_with_sigmaz_wp_ZMAX_LOII_omegam0268}. Although both models show a continuously reduced $N_{\mathrm{cen}}$ towarding to the massive end, $N_{\mathrm{cen}}$ in the Avila model exhibits a faster decay after the peak. This difference might imply that the AGN feedback mechanism in the SAM \citep{2018MNRAS.474.4024G} is too strong, resulting in quick quenching of galaxies at the massive end. As for $N_{\mathrm{sat}}$,  a power-law form can indeed reasonably describe the $N_{\mathrm{sat}}$ of ELGs under the current data. 

From the above comparison of the three models, we can conclude that since ELGs are mainly the central galaxies with small stellar mass, the form of $N_{\mathrm{cen}}$ at large stellar (halo) mass cannot be well constrained with the clustering data of ELGs only. Our results indicate that the following form can better describe the HOD of ELGs,
\begin{eqnarray}
\begin{aligned}
N_{\mathrm{cen}} \left(M\right) &= N^{\mathrm{Exp}}_{\mathrm{cen,Geach}} \left(M\right) + N^{\mathrm{Erf}}_{\mathrm{cen,Geach}} \left(M\right) \times \left(1+\frac{M}{M_\mathrm{c}}\right)^{\beta_{\mathrm{c}}} \\
&= F^{\mathrm{B}}_{\mathrm{c}}\left(1-F^{\mathrm{A}}_{\mathrm{c}}\right)\exp\left[-\frac{\log\left(M/M_\mathrm{c}\right)^2}{2\sigma^2_{\log M}}\right] \\
& +F^{\mathrm{A}}_{\mathrm{c}}\left[1+\mathrm{erf}\left(\frac{\log\left(M/M_\mathrm{c}\right)}{\sigma_{\log M}}\right)\right] \times \left(1+\frac{M}{M_\mathrm{c}}\right)^{\beta_{\mathrm{c}}} \\
N_{\mathrm{sat}} \left(M\right) &= N_{\mathrm{sat,Geach}} \left(M\right) \\
&= F_{\mathrm{s}}\left[1+\mathrm{erf}\left(\frac{\log\left(M/M_{\mathrm{min}}\right)}{\delta_{\log M}}\right)\right] \left( \frac{M}{M_{\mathrm{min}}} \right)^{\alpha_{\mathrm{s}}},
\end{aligned}\label{equ:newform}
\end{eqnarray}
where $\beta_{\mathrm{c}}$ characterizes the decay of $N_{\mathrm{cen}}$ at the high-mass end. This HOD preserves most features of the Geach HOD form, but  remedies it by introducing a gradually declining function at large mass. On the other hand, the behavior of  $N_{\mathrm{cen}}$ at large stellar (halo) mass can serve as a sensitive test for feedback models of galaxy formation. 

To check the performance of our modified Geach form (Equation \ref{equ:newform}), we compare it with our derived HODs of the four $L_\oiirm$ -selected subsamples in Figure \ref{fig:plot_modified_HOD_9p_compare} (the same as those shown in Figure \ref{fig:bestfittingHOD_logM50_compare_sim_cross_randomSM_SameCS_9SM_table_with_sigmaz_wp_ZMAX_LOII_omegam0268}, but shown as hollow circles). We assume that the data points of the derived HODs are equally weighted (assuming 10\% error), and fit them with Equation \ref{equ:newform}. The corresponding best-fitting results of the modified HOD model are displayed as solid lines. We can see that this HOD model can accurately describe the ELG occupation numbers at all halo mass and at all \oii luminosity. The parameters of this HOD model are listed in Table \ref{tab:modified_hod}.

\begin{deluxetable*}{cccccccccc}
	\tablenum{4}
	\tablecaption{The parameters of the modified Geach form for the four subsamples. \label{tab:modified_hod}}
	\tablehead{ \colhead{Name} & 
		\colhead{$\log M_{\mathrm{c}} \, [M_{\odot}\,h^{-1}]$} & \colhead{$\sigma_{\log M}$ } & \colhead{$F^{\mathrm{A}}_{\mathrm{c}}$} & \colhead{$F^{\mathrm{B}}_{\mathrm{c}}$} & \colhead{$\beta_{\mathrm{c}}$} & \colhead{$\log M_{\mathrm{min}} \, [M_{\odot}\,h^{-1}]$} & \colhead{$F_{\mathrm{s}}$} & \colhead{$\delta_{\log M}$ } & \colhead{$\alpha_{\mathrm{s}}$} 
	}
	\startdata
	$L0$&11.234&0.206&0.133&0.010&-0.185&11.690&0.015&0.516&0.947 \\
	$L1$&11.415&0.224&0.091&0.146&-0.187&11.668&0.012&0.516&0.939 \\
	$L2$&11.528&0.241&0.035&0.075&-0.168&11.723&0.005&0.508&0.940 \\
	$L3$&11.558&0.217&0.010&0.021&-0.065&11.783&0.001&0.492&0.950 \\ 
	\enddata
\end{deluxetable*}

\section{Summary} \label{sec:summary}
In this work, we constrain the ELG-halo connection using the auto and cross correlation functions of the galaxy subsamples from VIPERS. Combining the SHMR and ELG-stellar mass distribution, we provide a novel method to populate ELGs in cosmological simulations. Our main results are summarized as follows.

\begin{enumerate}
	\item Using the galaxy catalog from VIPERS, we construct four stellar mass-selected subsamples and four \oii luminosity-selected subsamples. We also take into account the redshift measurement uncertainty in our N-body simulation to make a fair comparison with the observations.
	\item Both the angular and radial selection functions of VIPERS have been carefully corrected. Particularly, to account for the radial selection effects caused by the $i$-band limit and the color-color cut, we adopt the $V_{\mathrm{max}}$ technique to generate the redshift distribution for the random samples. For all the galaxy subsamples, we measure the projected cross (auto) correlation functions in real-space and the multiple moments in redshift-space.
	\item To determine the SHMR, we apply the AM model proposed by \citep{2010MNRAS.402.1796W} to our N-body simulation. The theoretical cross (auto) correlation functions of different $\M$-selected subsamples are calculated by the tabulated method. We perform an MCMC analysis to explore the parameters space of SHMR. Our best-fitting SHMR can recover the observational correlation functions well. 
	\item We measure the ELG fractions $F_L\left(\M\right)$ as a function of stellar mass in the four $L_{\oiirm}$-selected subsamples. We demonstrate that the clustering of ELGs can be well matched both in the real-space and in the redshift-space if we use the above SHMR to assign stellar mass to (sub)halos and then randomly select the ELGs according to their fractions $F_L\left(\M\right)$ at a given stellar mass, as long as the satellite fraction $f_{\mathrm{sat}}$ is properly reduced. The method can be applied to constructing mock samples for ongoing and future redshift surveys, such as DESI, PFS and Euclid.
	\item We also derive the halo occupation numbers for the four ELG subsamples, and compare them with some of the previous HOD studies for ELGs. Our results indicate that the Geach form describes well the number of central galaxies at small and typical halo mass, but its constant form overpredicts the number at high halo mass. We propose a modified form,  Equation \ref{equ:newform}, for describing HOD of ELGs. The behavior at the high-mass reflects the feedback processes in galaxy formation. In addition, the power law form generally describes well the HOD of satellite galaxies.
	 
\end{enumerate}

In short, the cross correlations between ELGs and normal galaxies can play a significant role in constraining the ELG-halo connection. It is worth mentioning that our method can be combined with Photometric objects Around Cosmic webs (PAC) method \citep{2021arXiv210911738X}, which utilizes the cross correlation between a special spectroscopic sample (e.g., LRGs, QSOs) and a deep photometric sample, and thus can accurately measure the SHMR (SMF) in a wide stellar mass range. For the ongoing and future spectroscopic surveys, after the SHMR is determined using PAC, our method can be further developed and tested, and will provide a novel way to create \oii ELGs mock catalogs.
  
\acknowledgments

H.Y.G thanks Xiaokai Chen and Haojie Xu for their kind help. The work is supported by NSFC (12133006, 11890691, 11621303) and by 111 project No. B20019. We gratefully acknowledge the support of the Key Laboratory for Particle Physics, Astrophysics and Cosmology, Ministry of Education. This work made use of the Gravity Supercomputer at the Department of Astronomy, Shanghai Jiao Tong University.

This paper uses data from the VIMOS Public Extragalactic Redshift Survey (VIPERS). VIPERS has been
performed using the ESO Very Large Telescope, under the "Large Programme" 182.A-0886. The participating institutions and funding agencies are listed at
http://vipers.inaf.it. 
Based on observations collected at the European Southern Observatory, Cerro Paranal, Chile, using the Very Large Telescope under
programs 182.A-0886 and partly 070.A-9007. Also based on observations obtained with MegaPrime/MegaCam, a joint project of CFHT
and CEA/DAPNIA, at the Canada-France-Hawaii Telescope (CFHT),
which is operated by the National Research Council (NRC) of Canada,
the Institut National des Sciences de l’Univers of the Centre National
de la Recherche Scientifique (CNRS) of France, and the University of
Hawaii. This work is based in part on data products produced at TERAPIX and the Canadian Astronomy Data Centre as part of the CanadaFrance-Hawaii Telescope Legacy Survey, a collaborative project of
NRC and CNRS. This research uses data from the VIMOS VLT Deep Survey, obtained from the VVDS database operated by Cesam, Laboratoire d'Astrophysique de Marseille, France.

\software{Numpy \citep{5725236}, Scipy \citep{4160250}, Matplotlib \citep{4160265}, Astropy \citep{2013A&A...558A..33A}, scikit-learn \citep{scikit-learn}, emcee \citep{2013PASP..125..306F}}

\appendix
\begin{figure}
	\centering
	\includegraphics[scale=0.7]{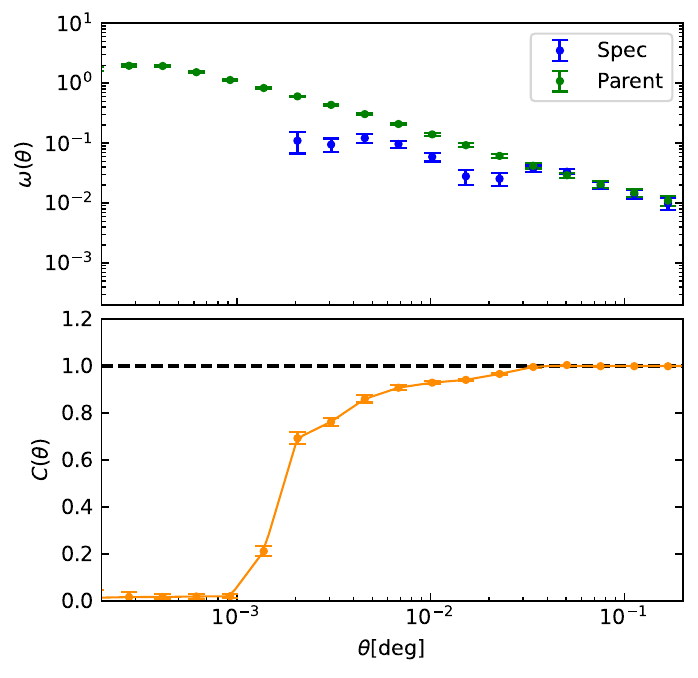}
	\caption{The angular completeness function  of VIPERS. In the top panel, the data points with error bar show the angular correlation functions of parent $w_{\mathrm{p}}\left(\theta\right)$ (green) and spectroscopic $w_{\mathrm{s}}\left(\theta\right)$ (blue) samples measured from 153 VIPERS mock catalog. The angular completeness function defined as $C\left(\theta\right)=\left[1+w_{\mathrm{s}}\left(\theta\right)\right]/\left[1+w_{\mathrm{p}}\left(\theta\right)\right]$ is plotted as orange points in the bottom panel. The orange solid line represents the linear interpolation of $C\left(\theta\right)-\log \theta$ relation.
		\label{fig:mock_angular_W1W4_interpolate_in_log}}
\end{figure}

\section{Correction to the small-scale clustering}
\label{sec:slit collisions}
Using the 153 VIPERS mock samples provided by \cite{2017A&A...604A..33P}, we compute the angular correlation function $w_{\mathrm{p}}\left(\theta\right)$ for parent photometric galaxies and $w_{\mathrm{s}}\left(\theta\right)$ for spectroscopic galaxies which is obtained by applying the silt assign algorithm to the parent catalog. In Figure \ref{fig:mock_angular_W1W4_interpolate_in_log}, we show the measurements of $w_{\mathrm{p}}\left(\theta\right)$ ($w_{\mathrm{s}}\left(\theta\right)$) in the top panel and $C\left(\theta\right)=\left[1+w_{\mathrm{s}}\left(\theta\right)\right]/\left[1+w_{\mathrm{p}}\left(\theta\right)\right]$ in the bottom panel. The two turning points of $C\left(\theta\right)$ clearly reflect the two typical scales that affect the small-scale clustering. We linearly interpolate the $C\left(\theta\right)-\log \theta$ relation and define the angular weight as $w^{\mathrm{A}}\left(\theta\right)=1/C\left(\theta\right)$.

\begin{figure}
	\vspace{0.1cm}
	\includegraphics[scale=0.75]{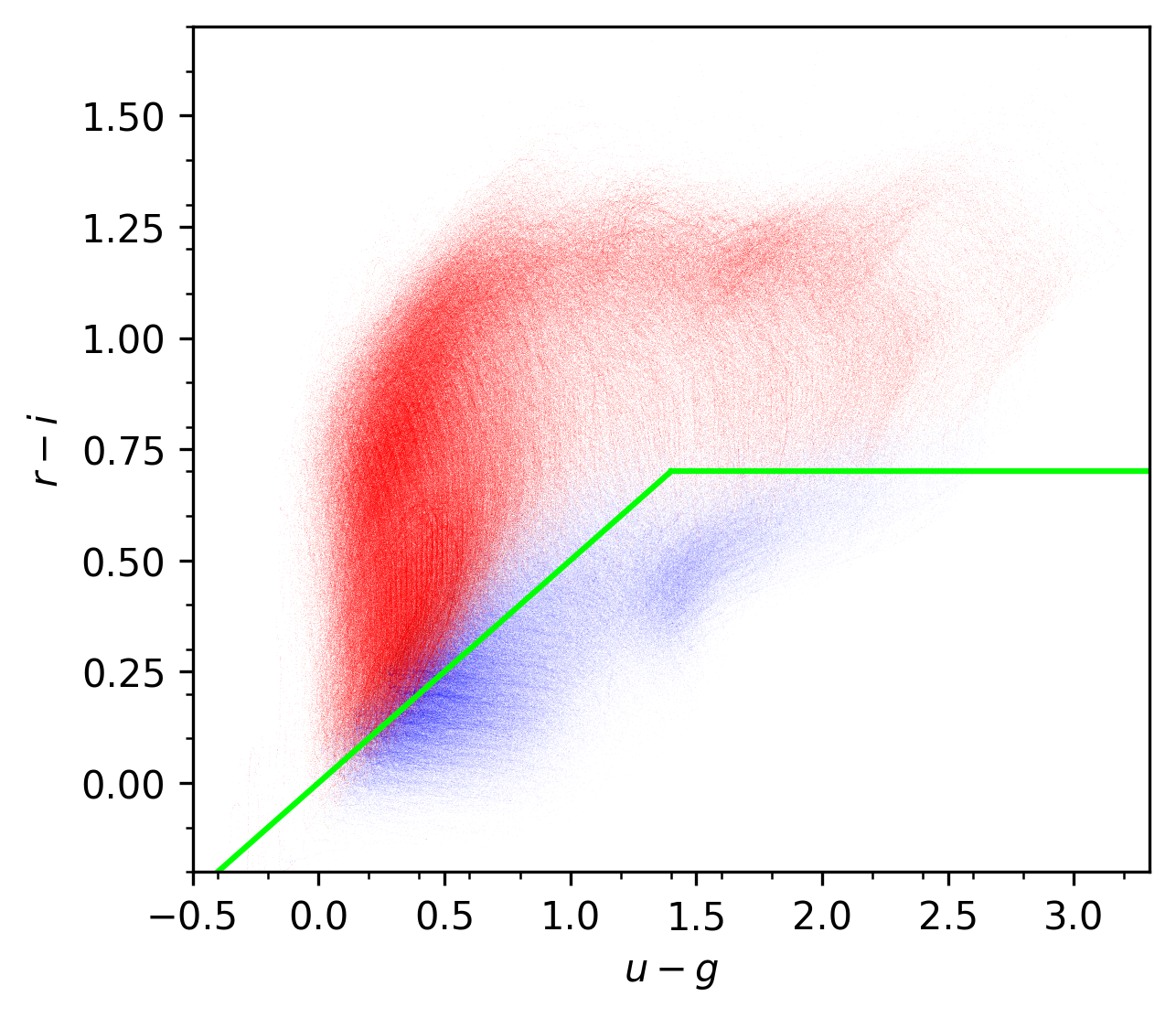}
	\caption{The color space distributions of the random points generated for the full galaxy sample. The blue (red) points denote the random points with $z<0.5$ ($z>0.5$). The lime solid line represents the VIPERS color-color cut (Equation \ref{equ:color-cut}) used to exclude the galaxies with $z<0.5$.
		\label{fig:color_check_ZMAX_weighted}}
\end{figure}

\begin{figure*}
	\centering
	\includegraphics[scale=0.4]{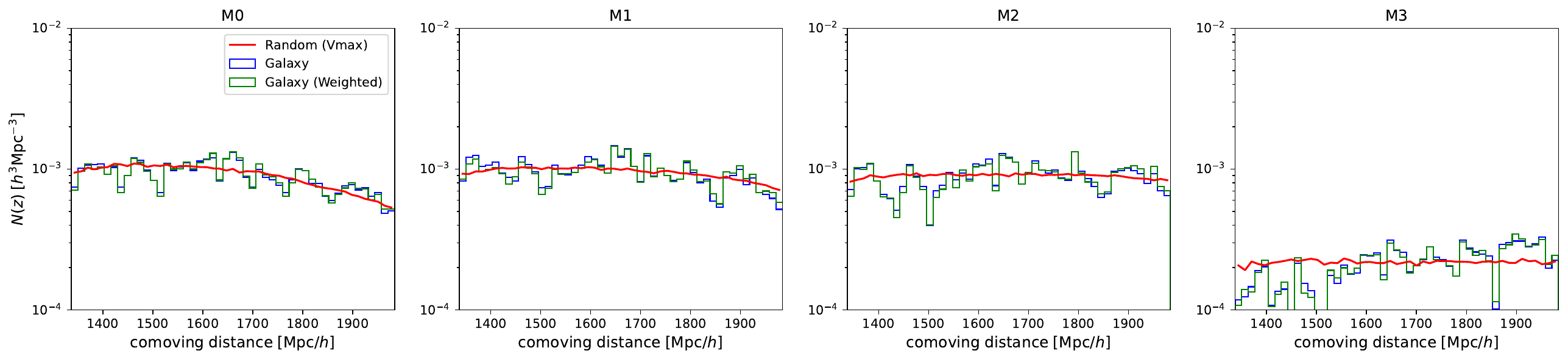}
	\caption{The radial distributions of the $\M$-selected subsamples. Similar to Figure \ref{fig:redshift_distribution}, the un-weighted (weighted) number densities as function of comoving distance for different subsamples are shown as blue (green) histograms. The radial distributions of corresponding random subsamples are plotted as red solid lines.  
		\label{fig:redshift_distribution_subsampleSM_ZMAX}}
\end{figure*}

\begin{figure*}
	\centering
	\includegraphics[scale=0.4]{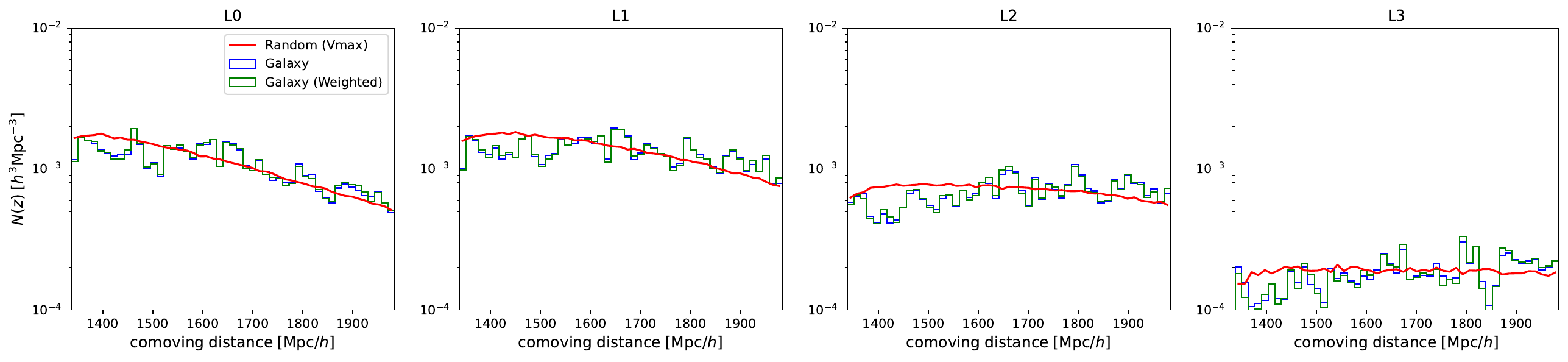}
	\caption{Similar to Figure \ref{fig:redshift_distribution_subsampleSM_ZMAX}, the radial distributions of the $L_\oiirm$ -selected subsamples. 
		\label{fig:redshift_distribution_subsampleLOII_evolution_cut_ZMAX}}
\end{figure*}

\section{Generating the redshift distribution for random sample}
\label{sec:vmax}
The $V_{\mathrm{max}}$ method \citep{2011MNRAS.416..739C, 2013A&A...557A..54D, 2017A&A...604A..33P, 2017A&A...608A..44D, 2020RAA....20...54Y} is adopted to produce the radial distribution for random samples. It is in principle much better than the method of randomly shuffling the observed redshifts in generating a random sample.  For each galaxy, we convert its $z_{\mathrm{max}}$ output by {\tt\string LE PHARE} \citep{2002MNRAS.329..355A,2006A&A...457..841I} into $V_{\mathrm{max}}$, which represents the maximum volume of this galaxy that can be observed in VIPERS. Consider that the SSR and TSR have slight impact on the redshift distribution of sample \citep{2017A&A...604A..33P}, we weight each galaxy by $w^{\mathrm{SSR}}\times w^{\mathrm{TSR}}$. Then we randomly select $N_{\mathrm{ran}}$ galaxies based on their probabilities (the probability of selecting different galaxies is not equal due to the weight). For each selected galaxy, we can generate a random point uniformly distributed in its $V_{\mathrm{max}}$ and convert the volume $V_{\mathrm{ran}}$ of this random point into its redshift $z_{\mathrm{ran}}$. As a result, the original best-fitting SED of the galaxy should be shifted to $z_{\mathrm{ran}}$ as a new SED of this random point. We can calculate the $u,g,r,i,z$ magnitudes for this random point and apply the color-color cut (Equation \ref{equ:color-cut}) to it. The Figure \ref{fig:color_check_ZMAX_weighted} presents the color space distributions of the random points generated by the above method for the full galaxy sample. We note that the color cut (lime solid line) can clearly distinguish random samples with redshifts lower than (blue points) and higher than (red points) 0.5. The radial distributions of random samples generated by this non-parametric method are shown in Figure \ref{fig:redshift_distribution} for the total sample and in Figure \ref{fig:redshift_distribution_subsampleSM_ZMAX}, and \ref{fig:redshift_distribution_subsampleLOII_evolution_cut_ZMAX} for the subsamples.

\bibliography{elg}{}
\bibliographystyle{aasjournal}

\end{document}